\documentclass{aa}  

\usepackage{graphicx}
\usepackage{txfonts}
\usepackage{lipsum}
\usepackage{subcaption}         % necessary for continued figures, example in section 3
\usepackage{lscape}             % to rotate a single page table, example in appendix.
\usepackage{placeins}           % useful with \FloatBarrier, to keep 
\usepackage[colorlinks=true, linkcolor=blue, citecolor=blue, urlcolor=blue]{hyperref}
\begin{document}

   \title{Dust evolution across cosmic times as seen through DUSTY-GAEA}

\authorrunning{O. Osman et al.}
%%%%%%%%%%%%%%%%%%%%%%%%%%%%%%%%%%%%%%%%
% Please do not include ORCIDs next to author names.
% Only ORCIDs authenticated by individual authors in EDP Sciences editorial system will be taken into account.
% ORCIDs included here will be removed.
%%%%%%%%%%%%%%%%%%%%%%%%%%%%%%%%%%%%%%%%

   \author{Omima Osman\inst{1, 2}\fnmsep\thanks{E-mail: omima.osman@inaf.it}, 
   Gabriella De Lucia\inst{1, 3},
    Fabio Fontanot\inst{1, 3},
    Lizhi Xie\inst{4},
    Michaela Hirschmann\inst{1, 5}}

   \institute{INAF -- Astronomical Observatory of Trieste, via G.B. Tiepolo 11, I-34143 Trieste, Italy
   \and  University of Khartoum -- Department of Physics, Al--Gamaa Ave, 11115 Khartoum, Sudan
   \and IFPU -- Institute for Fundamental Physics of the Universe, via Beirut 2, I-34151 Trieste, Italy
   \and Tianjin Astrophysics Center, Tianjin Normal University, Binshuixidao 393, 300384 Tianjin, China
    \and Institute for Physics, Laboratory for Galaxy Evolution, EPFL, Observatoire de Sauverny, Chemin Pegasi 51, 1290 Versoix, Switzerland}

   \date{Received November xx, 2025}

  \abstract{
 For many decades, dust has been recognised as an important ingredient in galaxy formation and evolution. This paper presents a novel self-consistent implementation of dust formation by stars, destruction by supernova shocks and hot gas, and growth within the dense interstellar medium (ISM) in the GAEA state-of-the-art galaxy formation model. Our new model, DUSTY-GAEA, reproduces well the dust buildup as a function of stellar mass out to $z\sim 6$, the scaling relations between the dust-to-gas/dust-to-metal ratios and stellar mass/metallicty in the local Universe, and the dust mass function in the local Universe and out to $z\sim 1$. In the framework of our model, dust growth dominates the cosmic dust budget out to $z\sim 8$, and we find that observational constraints beyond the local Universe can be reproduced only assuming such efficient dust growth in the dense ISM. Yet, reproducing the estimated number densities of dust-rich galaxies at higher redshifts remains challenging, as found also in independent theoretical work. We discuss our model predictions in comparison with both observational data and independent theoretical efforts, and highlight how further observational constraints at high redshifts would help constrain dust models. 
 }


\maketitle

%%%%%%%%%%%%%%%%%%%%%%%%%%%%%%%%%%%%%%%%%%%%%%%%%%%%%%%%%%%%%%%%%%%%%%%%%%%%%%%%%%%%%%%%%%%%%%%%%%%%%%%%%%%%%%%%%%%%%%%%%%%%%%%%%%%%%%%%%%%%%%%%%%%%%%%%%%%%%%%%%%%%%%%%%
\section{Introduction}

Interstellar dust is a challenging baryonic component  to study because of its complex interactions with both the interstellar medium (ISM) and stellar radiation. Dust provides a surface where numerous chemical reactions take place, leading to the formation of various molecules, including molecular hydrogen, essential for star formation (\citealp{Williams87}; \citealp{Cazaux04}; \citealp{Demyk11}; \citealp{Dulieu13}; \citealp{Bigiel08}; \citealp{Schruba11}). Moreover, collisions between dust grains and gas-phase species result in the depletion of gas-phase metals, cooling, and heating of the gas (\citealp{Black87}; \citealp{Draine11}; \citealp{Glover12}; \citealp{Klessen16}; \citealp{Zhukovska16}; \citealp{Zhukovska18}).

Dust grains also absorb and scatter stellar UV and optical radiation, re-emitting it in the infrared and submillimeter parts of the electromagnetic spectrum (\citealp{Draine03}; \citealp{Galliano18}; \citealp{Bianchi18}). Furthermore, far-ultraviolet (FUV) photons can eject energetic electrons from the surfaces of small dust grains (\citealp{Watson72}; \citealp{Draine78}; \citealp{Bakes94}; \citealp{Weingartner01}; \citealp{Wolfire03}; \citealp{Hill18}). These electrons are the main source of heating in the cold neutral medium (CNM) and diffuse atomic hydrogen (HI) regions (\citealp{Wolfire95}; \citealp{Ingalls02}). In summary, despite its small contribution to the mass budget of galaxies (only 0.1\% of the stellar mass; \citealp{Draine07}; \citealp{Smith12}), dust regulates the rates of chemical reactions that lead to the formation of different molecules in the ISM, chemical abundances, the dynamical and thermodynamical state of galaxies and their spectral energy distribution.

Our understanding of how dust forms and evolves has developed over decades of observational and theoretical studies. Dust is believed to form mainly in the ejecta of Asymptotic Giant Branch (AGB) stars and supernovae (SNe) type Ia and type II (\citealp{Dwek98}). SNe inject  into the ISM both oxygen-rich and carbon-rich dust (e.g. \citealp{Dwek98}; \citealp{Nozawa03}; \citealp{Sarangi18}), while AGB stars produce either oxygen-rich or carbon-rich dust, depending on the ratio of carbon to oxygen in their ejecta (\citealp{Whittet89}; \citealp{Sargent10}; \citealp{Srinivasan10}). Once injected into the ISM, dust grains undergo destruction processes driven by supernova remnants and interactions with the hot gas. These processes include sputtering, sublimation, and complete or partial vaporization (\citealp{Barlow78}; \citealp{Mckee89}; \citealp{Jones94}; \citealp{Nozawa06}; \citealp{Bianchi07}; \citealp{Yamasawa11}; \citealp{Andersen11}; \citealp{Bocchio16}). 

The dust grains that survive can eventually reach dense regions of the ISM, where they can grow through the accretion of gas-phase metals (\citealp{Dwek80}; \citealp{Dwek98}; \citealp{Hirashita13}). Depletion studies strongly support and constrain this mechanism of dust formation (\citealp{Duley78}; \citealp{Jones85}; \citealp{Savage96}). For example, \cite{Zhukovska16} and \cite{Zhukovska18} found that dust growth is essential to explain silicon and iron depletion levels in the Milky Way. Finally, dust grains could shatter or coagulate when they collide with each other, leading to changes in the grain size distribution (e.g. \citealp{Hirashita15}).

A large amount of information about dust abundance in galaxies and how it relates to their physical properties has been collected over the past decades. This information includes the scaling relations between dust mass and stellar mass (\citealp{Corbelli12}; \citealp{Santini14}; \citealp{Beeston18}; \citealp{Nersesian19}), dust mass and star formation rate (\citealp{daCunha10}; \citealp{Casey12}; \citealp{Santini14}; \citealp{Rowlands14}), the dust mass function of galaxies (\citealp{Dunne03}; \citealp{Vlahakis05}; \citealp{Eales09}; \citealp{Dunne11}; \citealp{Clemens13}; \citealp{Clark15}; \citealp{Beeston18}; \citealp{Pozzi20}), the dust-to-gas ratio versus metallicity (DtoG, \citealp{Issa90}; \citealp{Lisenfeld98}; \citealp{James02}; \citealp{Draine07b}; \citealp{Galametz11}; \citealp{Magrini11}; \citealp{Hirashita11}; \citealp{Hirashita13}; \citealp{Remy14}; \citealp{Relano18}), the dust-to-metal ratio versus metallicity (DtoM, \citealp{DeCia13}; \citealp{Zafar13}; \citealp{Sparre14}; \citealp{DeCia16}; \citealp{Wiseman17}; \citealp{Popping22}), and the gas fraction of galaxies versus dust mass (\citealp{Cortese12}). This wealth of data can be used to constrain dust models included in theoretical frameworks of galaxy formation.  

In the local Universe, dust produced by stars and  dust growth appear to play a dominant role in regulating the amount of dust observed and the scaling relations with other physical properties. At high redshift, the situation is less clear, as dust evolution may have proceeded quite differently. For instance, many galaxies appear to be over-abundant in  dust for their age, with dust masses equal to or exceeding 10$^7$ M$_\odot$ at z $>$ 4 (e.g., \citealp{Riechers14}; \citealp{Watson15}; \citealp{Pozzi20}), representing a challenge for  current theoretical models. This discrepancy raises questions about our understanding of the processes that control dust in the early stages of galaxy evolution. For example, it is unclear whether progenitors of AGB stars are effective dust producers at high redshift given the long time-scales involved  (see \citealp{Valiante09} for alternative view). It is also unclear whether dust growth continues to be a dominant dust formation channel at high redshift given that the gaseous metallicity decreases at earlier cosmic epochs. The effective stellar yields at high redshift are also unknown and could be significantly different from those measured in the local Universe. Observational data at these earlier cosmic epochs are also limited and potentially affected by systematics and selection effects that we probably still do not fully understand.

In recent years, theoretical models have been updated to include processes that account for the formation, destruction, and growth of dust (e.g.  \citealp{Bekki13}; \citealp{McKinnon16};  \citealp{Popping17}; \citealp{Vijayan19}; \citealp{Li19}; \citealp{Triani20}; \citealp{Yates24} ). A few models also include a treatment for the grain size distribution (e.g. \citealp{Hirashita15};  \citealp{Aoyama17};  \citealp{Hou17};  \citealp{Gjergo18};  \citealp{Granato21}; \citealp{Parente23}). These models reproduce reasonably well properties such as the local dust mass function, dust mass-stellar mass relation and dust-to-gas ratio versus metallicity relation in the local Universe. Observations at high redshift remain more challenging to reproduce. 

In this work, we introduce an updated version of the GAEA\footnote{\url{https://sites.google.com/inaf.it/gaea/home}} (GAlaxy Evolution and Assembly) model, that includes an explicit treatment for dust formation, destruction and growth. GAEA is a state-of-the-art semi-analytic model that successfully reproduces a wide range of observational results. Notably, GAEA employs an innovative approach for the non-instantaneous recycling approximation for the gas, metals, and energy by SNe and AGBs (\citealp{DeLucia14}), resulting in precise calculations of the ISM  metallicity history and providing an ideal framework for the implementation of dust physics. Semi-analytic models (SAMs) are particularly well-suited to address the challenges involved in dust prescriptions, as they allow the exploration of a vast parameter space across large cosmological volumes, while keeping computational costs reasonable. The paper is organized as follows: Section \ref{sec:sec2} presents our dust model and its variants. Results are detailed in Section \ref{results}. Our discussion and summary are presented in Sections \ref{Discussion} and \ref{summary}, respectively.

%%%%%%%%%%%%%%%%%%%%%%%%%%%%%%%%%%%%%%%%%%%%%%%%%%%%%%%%%%%%%%%%%%%%%%%%%%%%%%%%%%%%%%%%%%%%%%%%%%%%%%%%%%%%%%%%%%%%%%%%%%%%%%%%%%%%%%%%%%%%%%%%%%%%%%%%%%%%%%%%%%%%%%%%%

\section{DUSTY-GAEA}
\label{sec:sec2}

In this work, we use the Millennium (\citealp{Springel05}, MSI) and Millennium II (\citealp{Boylan09}, MSII) simulations. These model boxes with a side length of 500 and 100 Mpch$^{-1}$, respectively, and assume the following cosmological parameters: $\Omega_\Lambda$ = 0.75, $\Omega_m$ = 0.25, $\Omega_b$ = 0.045, n = 1, $\sigma_8$ = 0.9, and h = 0.73. The MSI contains 2160$^3$ particles and achieves a resolution of 8.61$\times$10$^8$ M$_\odot$, while MSII contains the same number of particles and achieves a resolution of 6.89$\times$10$^6$ M$_\odot$. The cosmological parameters adopted here differ from those from the Planck (\citealp{Planck16}) and WMAP9 (\citealp{Bennett13}) results. The impact on model predictions is expected to be minimal as demonstrated by \cite{Wang08} and \cite{Guo13}. Indeed, \cite{Fontanot25} showed that GAEA predictions converge well for runs using merger trees extracted from simulations at different resolutions and slightly different cosmological parameters, ranging from WMAP1 to Planck values.

Our starting reference model version is the one described in \cite{DeLucia24}. This model represents a significantly modified and extended version of the original model by \cite{DeLucia07}. In particular, it includes an updated parameterization of stellar feedback (\citealp{Hirschmann16}), a partition of cold gas into atomic and molecular components (\citealp{Xie17}), a gradual stripping of the hot and cold gas from satellite galaxies (\citealp{Xie20}), and an improved model for cold gas accretion onto black holes and explicit modeling of quasar driven winds (\citealp{Fontanot20}). Particularly relevant for this study is the innovative treatment of the non-instantaneous recycling of gas, energy, and metal  (\citealp{DeLucia14}). The model is calibrated to reproduce several observables, including the stellar mass function up to redshift $\sim$ 3, the HI and H$_2$ mass function in the local Universe, and the AGN bolometric luminosity function up to redshift $\sim$ 4 (see, e.g. \citealp{DeLucia24}). In the following we will provide a summary of the GAEA prescriptions mostly relevant for this work. For full details on these implementations, we refer the reader to the original papers. 

In summary, GAEA assumes that baryons are distributed in four different reservoirs: stars within galaxies, cold gas in galaxy discs (representing the ISM), hot gas associated with centrals and satellites, and ejected gas that has been reheated and expelled from halos by SNe- and AGN-driven winds. This ejected gas can later be reincorporated into the hot gas component (\citealp{Hirschmann16}; \citealp{Xie20}). Metals are always assumed to follow the circulation of baryons between the different reservoirs. For example, metals are locked into stars when they are formed, keeping the metallicity of the ISM constant. In the model, DUSTY-GAEA, presented in this work, we have introduced three additional dust reservoirs, corresponding to each of the gas reservoirs: dust in the cold gas, dust in the hot gas, and dust in the ejected gas.

As described in detail in \cite{DeLucia14}, chemical enrichment is modelled by projecting the information about the metals, gas and energy produced by each simple stellar population (SSP) in the future. This approach allows an accurate accounting of the timings and properties of the individual SSPs in model galaxies. The reference model adopts a Chabrier IMF (\citealp{Chabrier03})\footnote{The choice of the IMF affects the chemical enrichment at fixed yields since it alters the relative number of AGB stars, SNIa, and SNII. The reference model (GAEA) also includes a variant allowing for variable IMF prescriptions (see e.g. \citealp{Fontanot24}). However, in this work, we focus on the standard choice of a Universal, MW-like, IMF. We will address the effects of varying IMF on dust abundance at high redshift in our future works.}, and makes use of the stellar yield tables by \cite{Karakas10} for low- and intermediate-mass stars, \cite{Thielemann03} for SNeIa, and \cite{Chieffi04} for SNeII. The yields by \cite{Chieffi04} are defined for stellar masses between 13 and 35 M$_\odot$. Between 8 and 13 M$_\odot$, the yields are scaled proportionally to the stellar mass, while for more massive stars, the values corresponding to the highest stellar mass tabulated are used. In our version of the model, we have modified the code to follow explicitly the abundance of ten chemical elements (H, He, Mg, Si, O, C, Fe, S, Ca, Ti), and the chemical enrichment scheme such that all the metals are released into the cold gas regardless of the halo mass (see also \citealp{Cantarella25}).

\subsection{The dust model}

Taking advantage of the accurate approach to model chemical evolution included in GAEA, we implement an explicit model for the formation and evolution of dust. The dust model adopted consists of (i) dust formation in stellar winds of AGB stars, SNIa, and SNII, (ii) accretion of the ISM gas-phase metals on dust grains, i.e. dust growth, and (iii) dust destruction by SN remnants and sputtering in the hot gas. These processes are the main drivers of the dust mass evolution (\citealp{Hirashita13}) and their implementation is described in the following sub-sections.

\subsubsection{Dust formation by stars}

The total mass of the $j$th element ($j$ = C, O, Mg, Si, S, Ca, Fe, Ti) in dust from the $k$th type of stars (SNIa, SNII, and AGB stars) can be expressed as (\citealp{Dwek98}):
\begin{equation}
m_{\rm dust, \it j}^k= \delta_{\rm c, \it j}^k F_{\rm ej}(m_{\rm ej, \it j}^k),
\end{equation}
where $\delta_{\rm c, \it, j}^k$ is the fraction of the $j$th chemical element from the $k$th stellar type locked up in dust grains (i.e. the condensation efficiency), $F_{\rm ej}$ is a function determining the total mass of metals that can be used for dust formation, and $m_{\rm ej, j}^k$ is the mass of the $j$th element ejected from the $k$th stellar type.  

\subsubsection*{ $\ast$ Dust from AGB stars}

Depending on whether the stellar ejecta is carbon- or oxygen-rich, the dust produced by AGB stars would be either carbon- or oxygen-rich, respectively. This is because oxygen and carbon combine to form the maximum possible amount of CO, and as such either one of the two would be unavailable for dust formation by the end of the CO formation process (\citealp{Dwek98}). Following previous works, we consider the two cases described below.\\

Case 1: C/O $>$ 1, Carbon-rich dust

\begin{gather}
m_{\rm dust, \it C}^{AGB}= \delta_{\rm c, \it C}^{AGB} (m_{\rm ej, \it C}^{AGB} - 0.75m_{\rm ej, \it O}^{AGB}),  \\
m_{\rm dust, \it j}^{AGB} = 0,
\end{gather}
where j = O, Mg, Si, S, Ca, Fe, Ti.\\

Case 2: C/O $<$ 1, Oxygen-rich dust

\begin{gather}
m_{\rm dust, \it C}^{AGB} = 0,\\
m_{\rm dust, \it j}^{AGB}= \delta_{\rm c, \it j}^{AGB} m_{\rm ej, \it j}^{AGB}, \\
m_{\rm dust, \it O}^{AGB}= 16 \times \sum_j \delta_{\rm c, \it j}^{AGB} m_{\rm ej, \it j}^{AGB}/\mu_j,
\end{gather}
where j = Mg, Si, S, Ca, Fe, Ti and $\mu_j$ is the atomic mass of element $j$.

\subsubsection*{ $\ast$ Dust from SNIa and SNII}

SNe produce both carbon- and oxygen-rich dust since the macro-mixing of their ejecta allows the presence of carbon- and oxygen-rich regions (\citealp{Nozawa06}). We assume: \\
\begin{gather}
m_{\rm dust, \it j}^{SNIa/SNII}= \delta_{\rm c, \it j}^{SNIa/SNII} m_{\rm ej, \it j}^{SNIa/SNII}, \\
m_{\rm dust, \it C}^{SNIa/SNII}= \delta_{\rm c, \it C}^{SNIa/SNII} m_{\rm ej, \it C}^{SNIa/SNII}, \\
m_{\rm dust, \it O}^{SNIa/SNII}= 16 \times \sum_j \delta_{\rm c, \it j}^{SNIa/SNII} m_{\rm ej, \it j}^{SNIa/SNII}/\mu_j,
\end{gather}
where j = Mg, Si, S, Ca, Fe, Ti.\\

We adopt $\delta_{\rm c, \it j}^{SNIa/SNII}$ = 0.15 for all elements produced by SNIa and SNII and $\delta_{\rm c, \it j}^{AGB}$ = 0.2 for all elements produced by AGB stars (\citealp{Popping17}). These low condensation efficiencies account for inefficiency in dust production and dust destruction during ejection into the ISM by the reverse SN shock (\citealp{Popping17}; \citealp{Micelotta18}). Some observational and theoretical studies have shown that dust produced by SNIa is minimal (see e.g. \citealp{Gioannini17}; \citealp{Li19}; \citealp{Parente22}). Switching off their contribution, we also find a negligible contribution of SNIa to our predicted dust budget; as such, we keep their contribution in the present work. 

We have also tested the original \cite{Dwek98} condensation efficiencies, namely $\delta_{\rm c, \it j}^{SNIa/SNII}$ = 0.5 for carbon and 0.8 for all the other elements produced by SNIa and SNII, and $\delta_{\rm c, \it j}^{AGB}$ = 1. However, for the set of metal yields we adopt, the oxygen-rich dust produced adopting those efficiencies exceeds the available oxygen. We lowered the constant factor in Eqs. 6 and 9 from 16 to 10, following \cite{McKinnon16}, however, the problem was not resolved.

Adopting a constant condensation efficiency is a simplification of the formation process. Indeed, stellar evolution studies indicate varying dust condensation efficiencies with the initial mass of stars and metallicity (e.g. \citealp{Ferrarotti06}). This and the uncertainties in the stellar metal yields and the IMF could significantly influence the amount of dust produced by stars. We will study this in more detail in future work.

\begin{figure}
\centering
    \includegraphics[height=15cm,width=7.5cm]{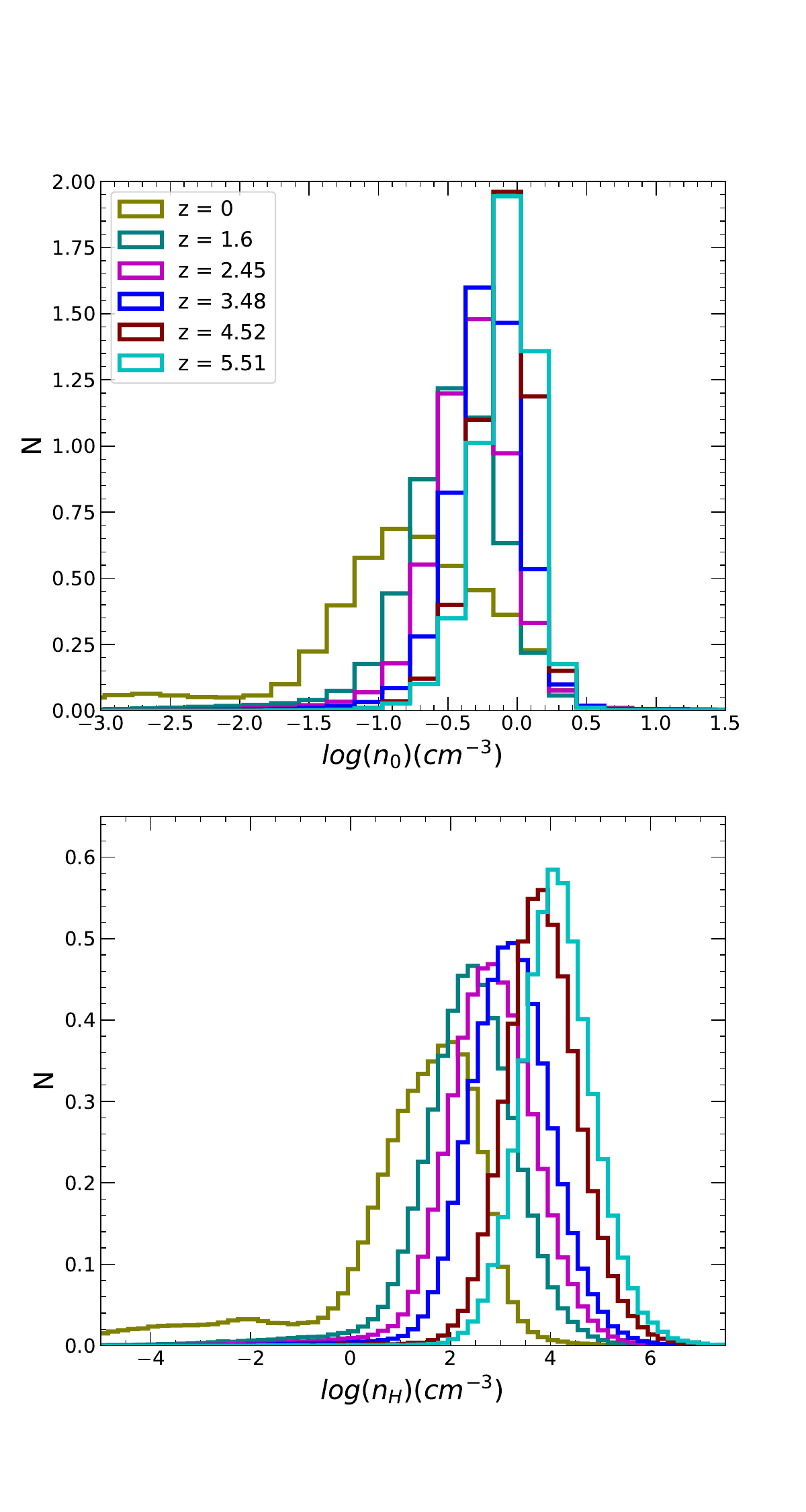}\par      
    \caption{Normalized histograms of the estimated density of the warm intercloud medium (top) and cold molecular gas (bottom) from z $\sim$ 0 to z $\sim$ 5.5. Predictions are based on the MSI merger trees. We limited the distributions to resolved galaxies (see Section \ref{results} for details).}
\vspace{-0.3cm}
\label{fig1}
\end{figure}

\subsubsection{Dust destruction}
\label{des}
Once dust is injected into the ISM, it experiences several destructive processes, including thermal and non-thermal sputtering (\citealp{Mckee89}; \citealp{Jones94}; \citealp{Jones04}; \citealp{Hirashita13}). Sputtering is caused by the high-velocity collisions between the gas phase atoms and ions with the dust grains. Thermal sputtering results from the random thermal motion of the gas species, as in the case of the hot gas in galaxy halos. Meanwhile, non-thermal sputtering arises from the relative motion between the gas and dust grains, as in the case of SNe forward shock waves. Metals removed from the dust grains are recycled to the ISM upon destruction. In the following, we describe our treatment of both processes. 

\subsubsection*{ $\ast$ Destruction by SNe forward shocks}
\label{SNeDes}

To model dust destruction by SNe, one needs to take into account a) the energy and evolution of the exploding SN, b) the properties of the ISM in which SN events occur  (i.e. density and metallicity), and c) the properties of the dust in this ISM (i.e. composition and grain size). Furthermore, ISM regions are often subject to several SNe explosions, and the total amount of dust destroyed in a region should be computed as the integral of the dust destroyed by all `overlapping' events. In practice, the problem is typically simplified by estimating the dust destroyed by a single event and then multiplying by the SNe rate corrected for SNe clustering.

We follow the approach described in \citet{Mckee89} to estimate the characteristic time scale of the process, the destruction time scale, $\tau_{des}$.
\begin{equation}
\tau_{des} = \begin{array}{rcl} \frac{M_{HI+HII}}{M_{swept} \gamma_{SN} \xi_{SN} \zeta_{SN}} & yr \end{array},
\label{des1}
\end{equation}
where
\begin{equation}
M_{swept} = \begin{array}{rcl} 6800(\frac{v_s}{100 kms^{-1}})(\frac{E_{SN}}{10^{51}erg}) & M_{\odot} \end{array},
\end{equation}
and
\begin{equation}
v_s = \begin{array}{rcl} 200(\frac{n_0}{1 cm^{-3}})^{\frac{1}{7}}(\frac{E_{SN}}{10^{51}erg}) & kms^{-1} \end{array}.
\end{equation}
In these equations, $M_{HI+HII}$ is the mass of the atomic and ionized gas, $M_{swept}$ is the mass of the ISM swept up by the shock wave, $\gamma_{SN}$ is SNe rate, $\xi_{SN}$ is the fraction of clustered SNe (= 0.36, \citealp{Mckee89}), and $\zeta_{SN}$ is a single SN destruction efficiency (= 0.34, \citealp{Nozawa06}). $v_s$ is the shock velocity, $E_{SN}$ is the energy of a single SN event, and $n_0$ is the ambient gas density. We estimate $\gamma_{SN}$ using the total energy released by SNIa and SNII events in each time step divided by the energy released in a single event ($\sim$ 10$^{51}$ ergs).

Most of the dust destruction is expected to occur in the warm intercloud medium (\citealp{Mckee89}); accordingly, we allow dust destruction only in this medium. Hence, we estimate n$_0$ using all the cold gas mass that is not in molecular form (for details about how the cold gas component of model galaxies is partitioned into different components, we refer the reader to the original paper by \citealp{Xie17}) and the volume, V$_0$, occupied by this gas. We compute V$_0$ as the volume of the galaxy disk multiplied by one minus the H$_2$ volume-filling factor (f$_{v,H_2} \sim$ 0.001, Table 1.3 in \citealp{Draine11}). Figure \ref{fig1} top panel shows n$_0$ distribution up to redshift 5.5 for the MSI runs (MSII runs yield similar results). The mass we used in the calculation is slightly overestimated since our calculation does not account for the presence of gas phases other than the warm ionized and the cold molecular. Therefore, we imposed an upper limit on the gas density consistent with what is reported in \cite{Draine11}, 1 cm$^{-3}$.\footnote{The single SN destruction efficiency, $\zeta_{SN}$, we adopted is consistent with this limit. The destruction efficiency increases quadratically with the ambient gas density according to \cite{Nozawa06} model (see their figure 10 and equation A3).} 

The same approach is used to estimate the density of the molecular gas ($n_H$, figure \ref{fig1} bottom panel), where V$_0$ is computed as the volume of the galaxy disk multiplied by H$_2$ volume-filling factor. It is worth noting that the H$_2$ volume-filling factor we adopted here is an estimation in the Milky Way and might not be the same in all galaxies at all redshifts. We adopt this approach trying to account for the redshift evolution of the ISM density instead of imposing a fixed density. 

The change in mass of the element $j$ in dust due to SNe destruction in the warm component of the cold gas is then estimated using the following equation from \cite{Dwek98}.

\begin{equation}
\Delta m_{dust, j}^{des}(t)= - \Delta t \,
m_{dust, j}(t) /\tau_{des},
\end{equation}
where $\Delta t $ is the integration time step and $m_{dust, j}(t)$ is the mass of the $j$th element already in dust. The same amount of metals removed from the dust phase, $\Delta m_{dust,j}^{des}$, is added to the metals in the (cold)gas phase.

\begin{equation}
\Delta m_{metal, j}(t) = \Delta m_{dust, j}^{des}(t).
\end{equation}

We also tested the model adopted by \cite{Popping17}, \cite{Vijayan19}, and \cite{Yates24}, based on the concept of the mass of the ISM cleared of dust, rather than the dust destruction efficiency. Furthermore, we tested the metallicity-dependent destruction model adopted by \cite{Triani20} and \cite{Parente23}, but still allowing the density of the gas to vary. Both models show similar destruction efficiency compared to our fiducial model (see Appendix \ref{appB} for further details).

\subsubsection*{ $\ast$ Destruction by thermal sputtering}

We implement thermal sputtering in both the hot and the ejected gas reservoirs following the prescriptions by \cite{McKinnon17} for the thermal sputtering time scale. This is based on the original formulation introduced  by \cite{Tsai95}:

\begin{equation}
\tau_{spu} = \begin{array}{rcl}  0.17 (\frac{\bar{a}}{0.1\mu}) (\frac{10^{-27}}{\rho}) [(\frac{2\times 10^6}{T})^{2.5} + 1] & Gyr \end{array},
\end{equation}
where $\rho$ is the gas density in g cm$^{-3}$ units, and T is the temperature in K. We used the virial temperature and assumed a homogeneous spherical density profile bounded by the virial radius to estimate the gas density. We further assume $\bar{a}$ = 0.1 $\mu$m for the average grain size, and that the ejected and hot reservoirs have the same properties. Accordingly, the change in mass of the element $j$ in dust due to sputtering in the hot or ejected gas is:

\begin{equation}
\Delta m_{dust, j}^{spu}(t)= - \Delta t \,
m_{dust, j}(t)/3\tau_{spu}.
\end{equation}

The same amount, $\Delta m_{dust,j}^{spu}$, is added to the (hot/ejected)gas phase metals.

\begin{table*}
\centering
\caption{Parameter values for the dust growth model based on \cite{Zhukovska08}}
\begin{tabular}{lllllll}
\hline
{Model}
& {Dust species}
& {Key element}
& {$\rho_{dust}$}
& {$A_{dust}$}
& {$A_{metals}$}
& {$b_{dust}$} \\
\hline
{Silicates} & Si or Mg$^a$ & 3.13& 121.41 & 24.3 or 28.1 & 1 or 1.06\\
{Carbon}  & C & 2.25 & 12.01 & 12.01 & 1\\
Metallic iron & Fe & 7.86 & 56 & 56 & 1\\
Silicon carbide$^b$ & Si & -- & -- & -- & --\\
\hline
\end{tabular}\\
$^a$The element with the least abundance is considered the key element among the two.
$^b$Inferred using the amount of Carbon bound in Silicon carbide and according to the formula: $\Sigma_{C, SiC} = \frac{12}{40}\Sigma_{Si}$, where $\Sigma_{Si}$ is the Si abundance after the formation of Silicates (\citealp{Zhukovska08}). 

\label{table2}
\end{table*}

\subsubsection{Dust growth in dense media}

Dust destruction by SNe is a rather efficient process, where only about 10 to 20 per cent of the dust survives  (\citealp{Mckee89}; \citealp{Dwek98}). The remaining dust could hardly account for the vast amounts of dust observed in the local and distant Universe (\citealp{Jones94}; \citealp{Bertoldi03}; \citealp{Mattsson11}; \citealp{Kuo12}; \citealp{McKinnon16}; \citealp{Ginolfi18}; \citealp{McKinnon18}; \citealp{Shivaei24}). To alleviate this tension, an efficient additional source of dust must be present. \cite{Dwek80} proposed dust growth via accretion of the gas phase metals in the dense metal-rich regions of the ISM as a viable process. Studies of ISM depletions strongly support and constrain this scenario (\citealp{Duley78}; \citealp{Jones85}; \citealp{Savage96}; \citealp{Zhukovska16}; \citealp{Zhukovska18}). Theoretical studies also support dust growth as a viable mechanism to explain observations (e.g. \citealp{Dwek98}; \citealp{Bekki13}; \citealp{Aoyama17}; \citealp{Popping17}).

Considering the uncertainties related to the modelling of this specific process, we adopt two frameworks to implement dust growth in our galaxy formation model. These are based on the formalisms introduced in \cite{Dwek98} and  \cite{Zhukovska08}. Both frameworks follow the evolution of the dust mass, omitting the evolution of its grain size distribution. However, unlike the Dwek framework, Zhukovska et al. take into account the dust grain composition, allowing us to have predictions of oxygen depletion of about $\sim$ 0.03 - 0.2, consistent with the depletion level adopted in emission-line modelling studies (e.g. \citealp{Groves04}; \citealp{Gutkin16}). 

\subsubsection*{ $\ast$ Dwek framework}

The growth time scale varies according to the ISM properties (i.e. density, temperature and metallicity) and the grain size. We use the following formula by \cite{Asano13}:

\begin{equation}
\label{Asano}
C_s \tau_{\rm acc, \it j} =  \begin{array}{rcl} 2\times 10^7  (\frac{\bar{a}}{0.1\mu})  (\frac{n_{H}}{100cm^{-3}})^{-1}  (\frac{T}{50K})^{-\frac{1}{2}}  (\frac{Z_j}{Z_\odot})^{-1} & yr \end{array},
\end{equation}
where $C_s$ (= 1) is the dust sticking coefficient, the probability that a metal atom or ion sticks to the dust grain after colliding, and $Z_j$ is the abundance ratio of the element $j$. We model the increase in dust mass due to the growth using a slightly modified form of the formula by \cite{Dwek98} (\citealp{Popping17}; \citealp{Triani20}):

\begin{equation}
\label{DWG}
\Delta m_{dust,j}^{\rm acc}(t)=\Delta t (1-f_{\rm dust, j}) 
m_{dust,j}(t) f_{H_2}/\tau_{\rm acc, j}.
\end{equation}

$f_{\rm dust, j}$ is the fraction of the $j$th element that is locked up in  dust and $f_{H_2}$ is the molecular hydrogen fraction as calculated in \cite{Xie17}. We adopt 0.1 $\mu$m and 50 K for the grain size and the cold gas temperature, respectively. We estimate the cold/molecular gas density following the same approach used to estimate the density for SNe destruction (see Section \ref{SNeDes} and figure \ref{fig1}). 

The dust growth process leads to a depletion of the gas phase elements; accordingly, the mass of the $j$th element in the gas decreases by $\Delta m_{dust,j}^{\rm acc}(t)$.

\subsubsection*{ $\ast$ Zhukovska et al. framework}

This framework also takes into account the recycling of the gas between the different phases in the ISM. It introduces an effective exchange time scale in which all the ISM is cycled through the molecular phase:

\begin{equation}
\tau_{exch, eff} = \frac{\tau_{exch\times(1-f_{H_2})}}{f_{H_2}}.
\end{equation}

$\tau_{exch}$ is the lifetime of molecular clouds (= 20 Myrs; \citealp{Murray10}). The increase in the dust mass due to growth is then estimated using the following equation.
\begin{equation}
\Delta m_{dust,j}^{\rm acc}(t)=\Delta t (f_{j, con}M_{metals,j}(t) - m_{dust,j}(t))/\tau_{\rm exch, eff},
\end{equation}
where
\begin{equation}
f_{j, con} = [(f_{j,0}(1+\frac{\tau_{exch}}{\tau_{acc}}))^{-2} + 1]^{\frac{-1}{2}},
\end{equation}
and
\begin{equation}
\tau_{acc} = \begin{array}{rcl} 46 \times \frac{b_{dust}A_{metals}^{\frac{1}{2}}}{A_{dust}}(\frac{\rho_{dust}}{3 gcm^{-3}})(\frac{3.5\times10^{-5}}{Z_j})(\frac{10^3cm^{-3}}{n_H}) & Myr \end{array}.
\label{zhu}
\end{equation}

$f_{j, con}$ is the condensation degree of a metal species $j$ at the end of the cloud lifetime, i.e. the mass fraction of the species $j$ condensed into dust, and $f_{j, 0}$ is the initial condensation degree.  $M_{metals,j}$ is the total mass of the $j$th element in the gas and dust. $\rho_{dust}$ is the dust grain bulk density, and $b_{dust}$, $A_{metals}$ and $A_{dust}$ are parameters determined by the dust grain composition. Following \cite{Zhukovska08}, we adopt carbon dust, silicates (Olivine and Pyroxene with Olivine fraction of 0.32), silicon carbide and metallic iron as dust grain species, hence we adopt the same parameter values adopted in the original paper (also shown in Table \ref{table2}. See also Section 4.3 of \citealp{Zhukovska08}). Equation \ref{zhu} is derived averaging with respect to the MRN grain size distribution (\citealp{Mathis77}).

%%%%%%%%%%%%%%%%%%%%%%%%%%%%%%%%%%%%%%%%%%%%%%%%%%%%%%%%%%%%%%%%%%%%%%%%%%%%%%%%%%%%%%%%%%%%%%%%%%%%%%%%%%%%%%%%%%%%%%%%%%%%%%%%%%%%%%%%%%%%%%%%%%%%%%%%%%%%%%%%%%%%%%%%%

\section{Results}
\label{results}

\begin{table*}
\centering
\caption{Main characteristics of the dust model presented in this work}
\begin{tabular}{llll}
\hline
{Model}
& {Formation}
& {Destruction}
& {Growth}\\
\hline
{DUSTY-GAEA (fiducial)}
 & AGB, SNe Ia, SNe II& SNe, sputtering in the hot gas  & Zhukovska et al. (2008) model\\
{DUSTY-GAEA-Dwek}  & AGB, SNe Ia, SNe II & SNe, sputtering in the hot gas & Dwek (1998) model\\
DUSTY-GAEA-NoG & AGB, SNe Ia, SNe II & SNe, sputtering in the hot gas & --\\
\hline
\end{tabular}
\vspace{-0.15cm}
\label{table1}
\end{table*}

\begin{figure*}
\centering
    \makebox[\textwidth][c]{\includegraphics[width=1.1\textwidth]{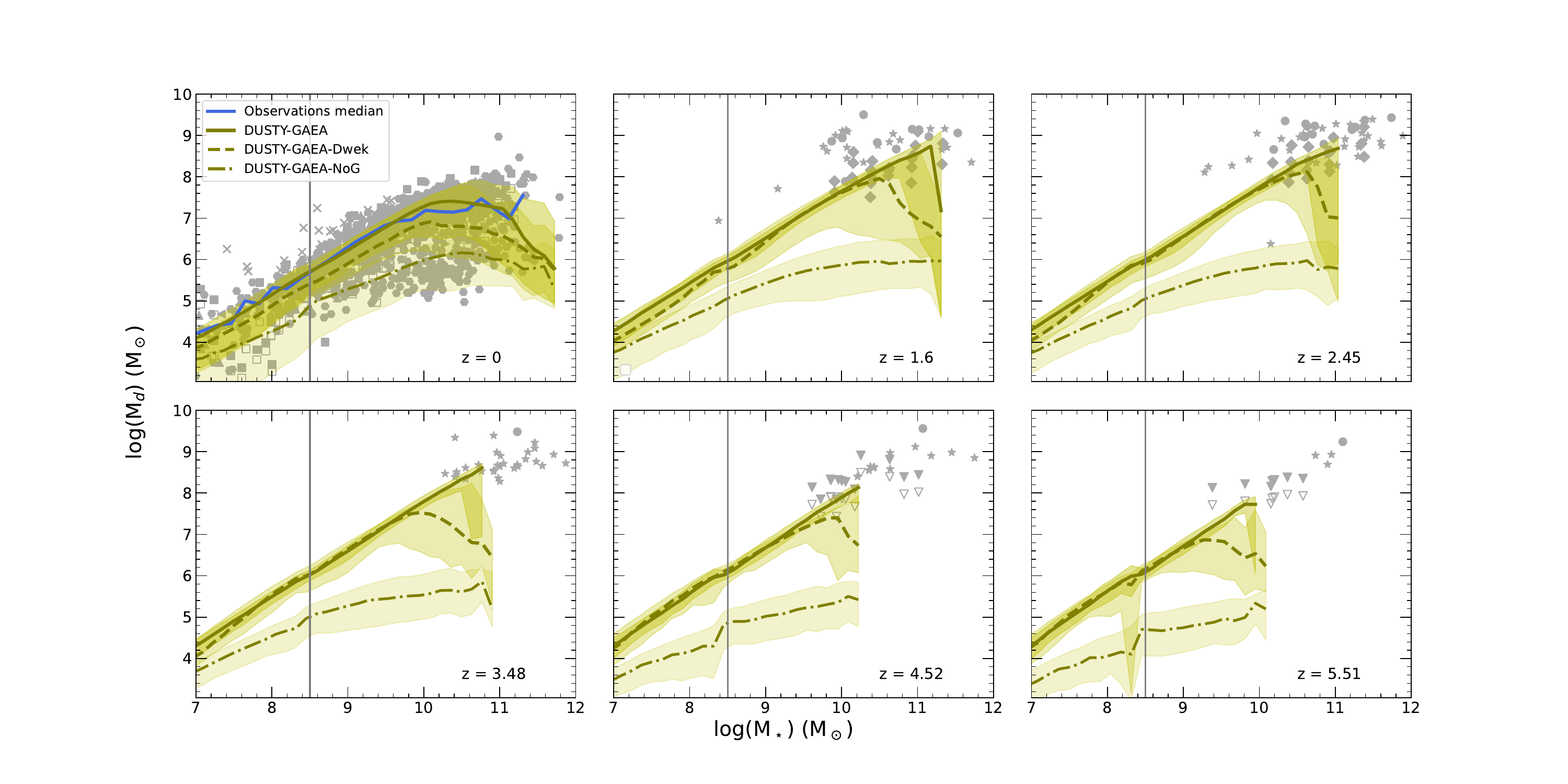}\par }     
    \caption{The dust mass as a function of the stellar mass at different redshifts. The solid olive lines represent the median predictions from our fiducial model (DUSTY-GAEA), while the dashed and dash-dotted lines correspond to predictions from DUSTY-GAEA-Dwek and DUSTY-GAEA-NoG, respectively. Shaded areas represent the 16th-84th percentile region. The solid vertical lines indicate the transition between predictions based on MSII and MSI. Symbols represent observational data from \cite{Clark15} (crosses), \cite{Remy15} (squares), \cite{Grossi15} (up triangles), \cite{DeVis19} (hexagons), \cite{Nersesian19} (right triangles), \cite{Beeston18} (left triangles), \cite{Santini14} (diamonds), \cite{Rowlands14} (circles), \cite{daCunha15} (stars), and \cite{Pozzi20} (down triangles).}
\vspace{-0.3cm}
\label{fig3}
\end{figure*}

In this section, we present the main dust scaling relations, the dust mass function and the dust cosmic evolution predicted by three variants of our dust model. As our fiducial model, we choose the model that adopts the Zhukovska et al. framework for dust growth. We will denote this model DUSTY-GAEA in the following. We will justify this choice further in the text. Along with DUSTY-GAEA, we discuss two other model variants: DUSTY-GAEA-Dwek, adopting the Dwek framework for dust growth, and a model where dust growth is switched off, DUSTY-GAEA-NoG. Table \ref{table1} presents the main characteristics of the three models. In our analysis, we include all galaxies above 10$^{8.5}$ M$_\odot$ from the Millennium I simulation (MSI) and galaxies between 10$^7$ and 10$^{8.5}$ M$_\odot$ from the Millennium II simulation (MSII). We follow \cite{Yates24} in combining predictions from MSI and MSII, weighting each galaxy by the inverse of the effective volume of the underlying N-body simulation. We note that discontinuities at the transition region between the two simulations are still visible in at least some scaling relations. 

\begin{figure*}
\centering
   \makebox[\textwidth][c]{ \includegraphics[width=1.1\textwidth]{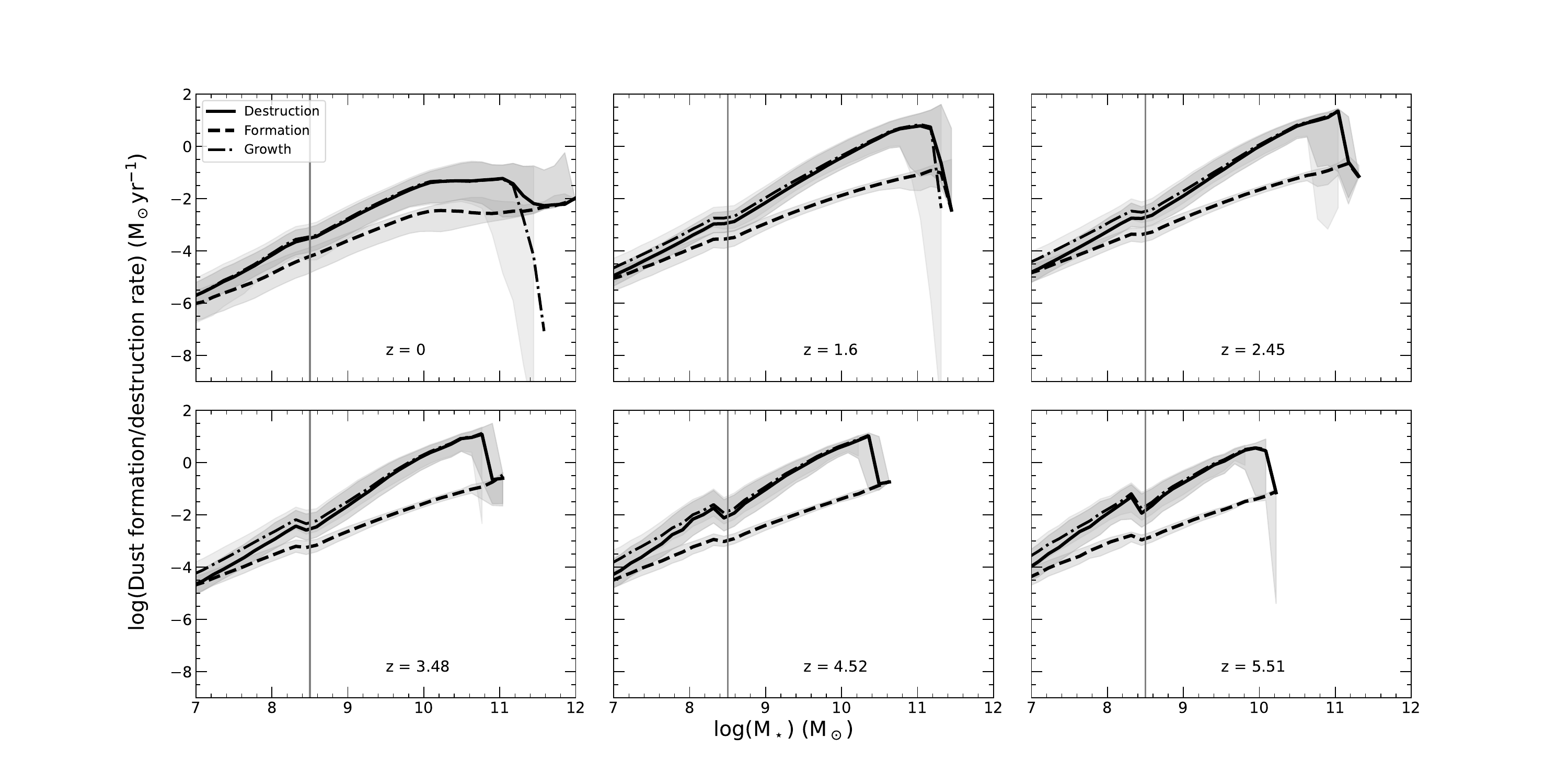}\par}      
    \caption{The dust formation, destruction and growth rates as predicted by the DUSTY-GAEA model. The solid-dashed, solid and dashed-dotted lines represent the dust formation rates by stars, destruction by SNe forward shocks, and growth in the dense ISM, receptively. Shaded areas represent the 16th-84th percentile region. The solid vertical lines indicate the transition between predictions based on MSII and MSI.}
\vspace{-0.3cm}
\label{rates}
\end{figure*}

\subsection{Dust properties versus stellar mass}
\label{dust-mass}
The dust mass-stellar mass relation from redshift 0 to $\sim$ 5.5 is shown in figure \ref{fig3}. The solid olive lines represent the median predictions from our fiducial model (DUSTY-GAEA), while the dashed and dash-dotted lines correspond to predictions from DUSTY-GAEA-Dwek and DUSTY-GAEA-NoG, respectively. Shaded areas represent the 16th-84th percentile region. Symbols denote observational data from various sources, including \cite{Clark15}, \cite{Remy15}, \cite{Grossi15}, \cite{DeVis19}, \cite{Nersesian19}, \cite{Beeston18}, and \cite{Santini14} at z $\sim$ 0; \cite{Rowlands14}, \cite{daCunha15}, and \cite{Santini14} for z between $\sim$ 1.6 and 2.45; and \cite{Rowlands14} and \cite{daCunha15} at higher redshifts. The ALPINE data from \cite{Pozzi20} are at redshifts $\sim$ 4.5 and 5.5. Filled symbols indicate dust mass estimates using a dust temperature of 25 K, while the empty ones indicate estimates using a temperature of 35 k. The blue solid line at z $\sim$ 0 represents the median of the observational data.

All three model variants predict an increase in dust mass as a function of stellar mass across all redshifts. At z $\sim$ 0, the relation bends down beyond a stellar mass of 10$^{10}$ M$_\odot$. Including only star-forming galaxies would remove this bend, indicating that this bend it is driven by the influence of passive galaxies in these high stellar mass bins. In these galaxies dust destruction by supernovae plays a significant role in reducing dust abundance. 

At higher redshifts, the model DUSTY-GAEA-Dwek shows similar bending, which is due to a higher destruction efficiency in this case. In contrast, the model DUSTY-GAEA does not exhibit such bending at higher redshifts. This is because the dust growth model in DUSTY-GAEA is more efficient and able to counterbalance destruction by supernovae up to high stellar masses (SNe rate increases with stellar mass and hence dust destruction). Generally, the different growth (and destruction) recipes we tested differ only slightly in their dust mass predictions for low-mass galaxies. At the high mass end, there may or may not be a bend in the relation depending on the specific assumptions. The model with no growth shows a flattening of the relationship above stellar mass of 10$^{8}$ M$_\odot$ across a larger stellar mass range. In this model, the production by stars is very efficient in increasing the dust mass at low stellar masses, where destruction is still relatively inefficient. As the stellar mass increases the two processes, formation and destruction, come to some sort of equilibrium resulting in the flattening. The gap between this model and the other models increases with stellar mass and redshift. 
At z $\sim$ 0, the trend of the observational data is well reproduced by the medians of the model variants DUSTY-GAEA and DUSTY-GAEA-Dwek. Beyond redshift zero, the DUSTY-GAEA variant performs better in reproducing observations up to z $\sim$ 5.5. The DUSTY-GAEA-Dwek model trend starts declining about 0.4-0.8 dex in stellar mass before the DUSTY-GAEA trend stops, limiting the model's ability to reproduce the dust mass of galaxies in the high stellar mass end. It is intriguing to note that these two model variants predict different dust abundances in these massive galaxies.

\begin{figure*}
\centering
    \makebox[\textwidth][c]{\includegraphics[width=1.1\linewidth]{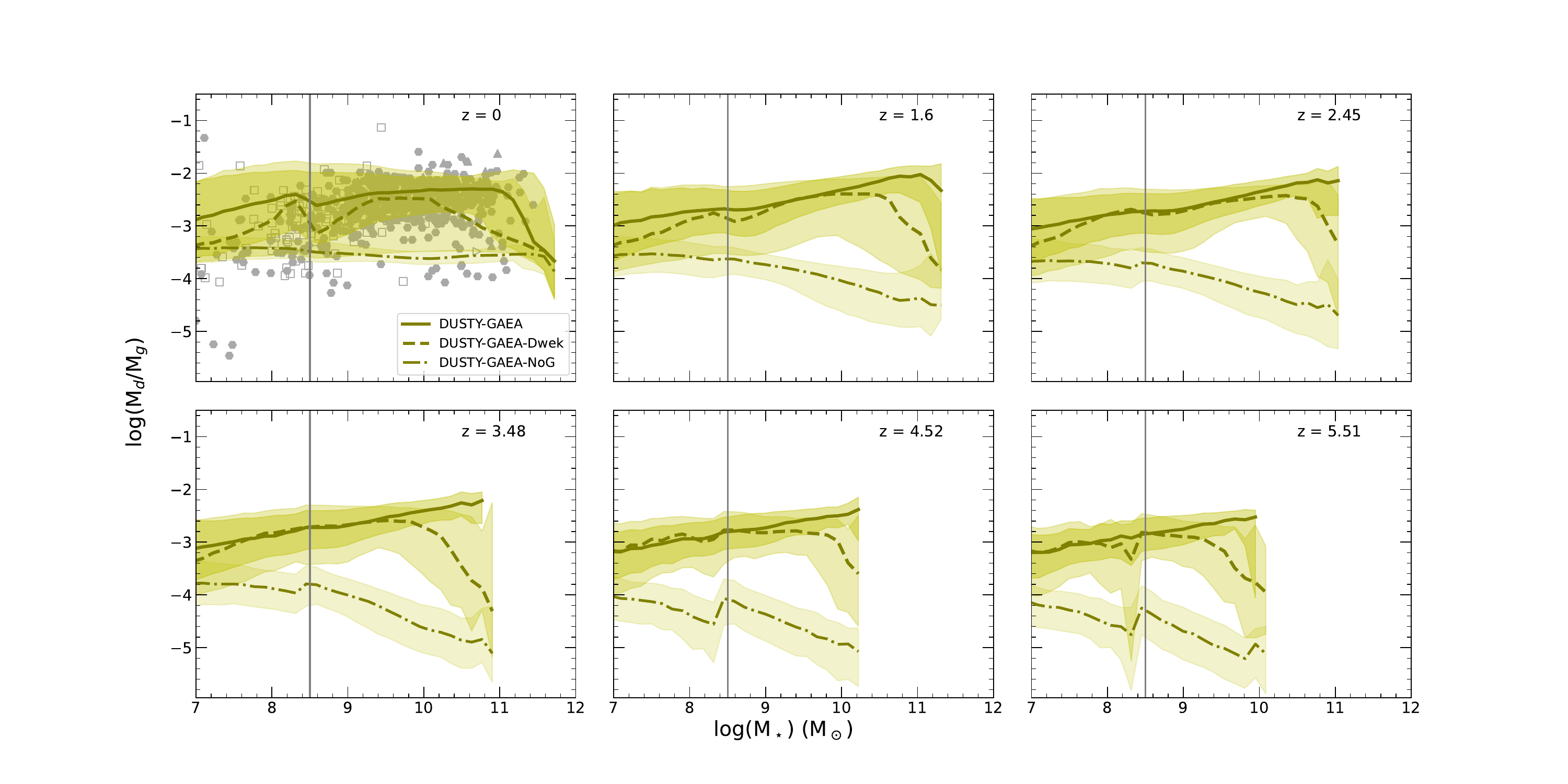}\par}      
    \caption{The dust-to-gas (DtoG) ratio as a function of the stellar mass at different redshifts. Line styles and shaded areas have the same meaning as  in figure \ref{fig3}. Symbols represent the observational data by \cite{Grossi15} (up triangles), \cite{Nersesian19} (right triangles), and \cite{DeVis19} (hexagons). Open symbols correspond to cases where only atomic hydrogen was used to estimate the total gas mass.}
\vspace{-0.3cm}
\label{fig4}
\end{figure*}

Dust abundance in galaxies is estimated through several methods, primarily falling within two main approaches. The first approach assumes that galaxies behave as optically thin sources with either single or double temperatures, where infrared spectral energy distributions of both local and high-redshift galaxies are typically modelled using 1- or 2-component modified blackbody. The second approach uses more complex dust models, such as those by \cite{Draine07}, that account for both a varying radiation field heating the dust and different dust compositions. Both approaches carry uncertainties, up to a factor of three (see \citealp{Popping17}; \citealp{Vijayan19}; \citealp{Triani20} for more details). The observational data presented here encompass both approaches and these methodological differences—as well as their associated uncertainties—should be kept in mind when compared to our model predictions. Furthermore, data from \cite{daCunha15}, \cite{Santini14}, and \cite{Rowlands14} are biased toward bright, relatively high star-forming galaxies, as these studies focus on submillimeter galaxies. Studies focused on Damped Lyman Alpha (DLA) systems like \cite{DeCia16}, \cite{Wiseman17}, and  \cite{Peroux20} indirectly infer dust properties at high-z using absorption lines of the gas phase species, adding another layer of systematic errors and uncertainties. Due to the diverse observations presented in this work and the diverse origins of errors and systematic errors carried by them, we did not attempt to convolve our model predictions with any observational errors.

It is evident from the comparison between the three model variants that dust growth is an important process for enriching galaxies at all redshifts but not necessarily at all stellar masses. To understand the relative importance of growth, we studied the dust formation, destruction and growth rates for each model. We show results from DUSTY-GAEA in figure \ref{rates}. The rates based on the DUSTY-GAEA-Dwek model version are very similar, and shown for completeness in  Appendix \ref{appA}. In figure \ref{rates}, solid-dashed, solid and dashed-dotted lines show the dust formation rates by stars, destruction by SNe forward shocks, and growth in the dense ISM, respectively. The rates of dust formation by stars and growth in the ISM are similar at low stellar masses, while they are about two orders of magnitude apart at the high stellar mass end, indicating an increasing efficiency of dust growth with stellar mass. The formation rate increases linearly with the stellar mass (a slope between 0.94 and 1.2 depending on the redshift), while the growth rate increases slightly more steeply with stellar mass (a slope between 1.34 and 1.46 depending on the redshift). The destruction rate follows closely the dominant formation process, which is growth. Rates in the DUSTY-GAEA-Dwek model behave similarly, except that the growth trends are a bit steeper (a slope between 1.30 and 1.73 depending on the redshift). In both models, rates of dust formation by stars could hardly exceed the growth in the ISM rates.

\begin{figure*}
\centering
    \makebox[\textwidth][c]{\includegraphics[width=1.1\linewidth]{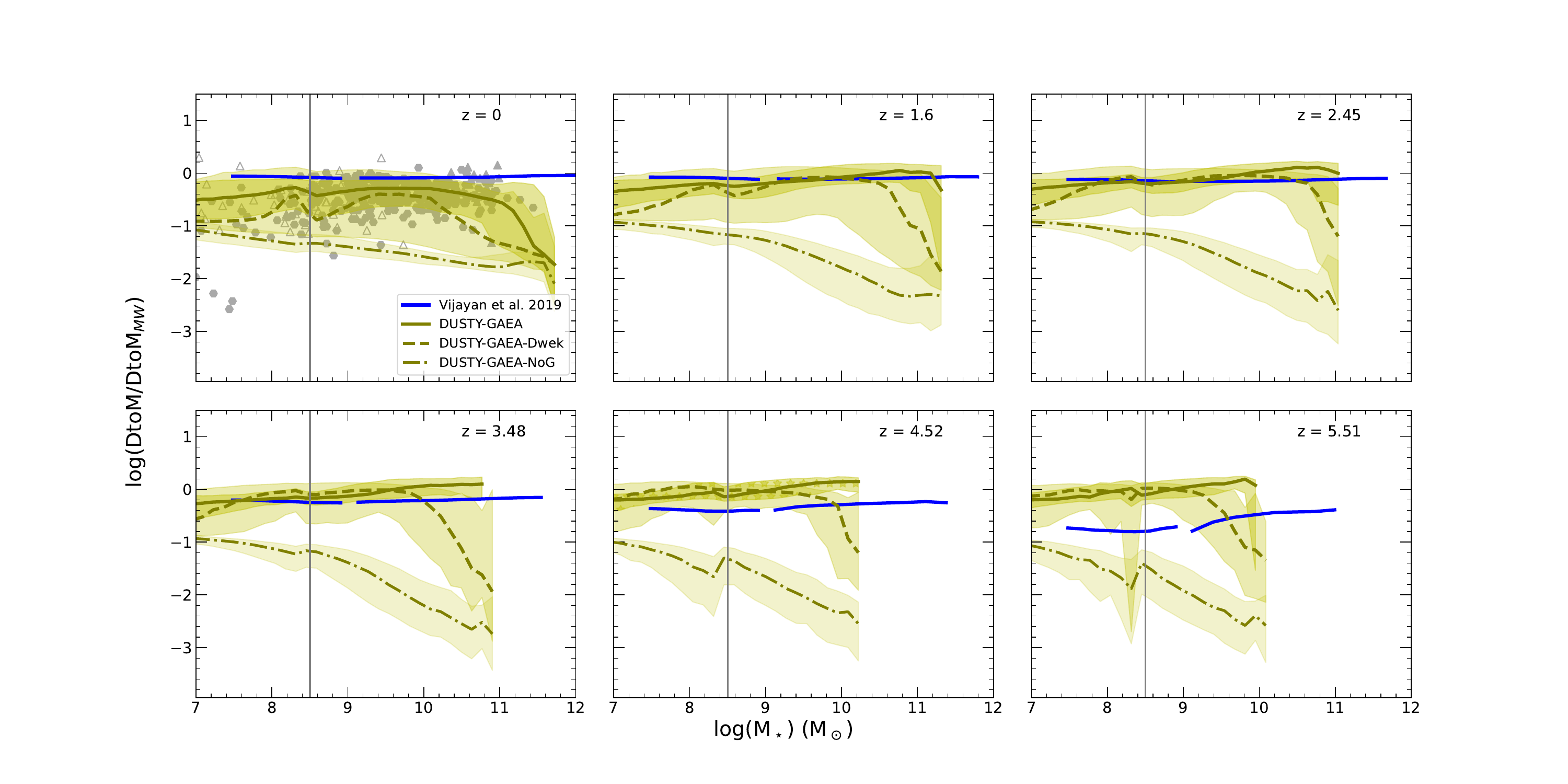}\par }     
    \caption{The dust-to-metal (DtoM) ratio normalized to the Milky Way value (i.e. 0.44) as a function of the stellar mass at different redshifts. Line styles and shaded areas have the same meaning as  in figure \ref{fig3}. Symbols represent the observational data by \cite{Grossi15} (up triangles) and \cite{DeVis19} (hexagons). Open symbols indicate cases where only atomic hydrogen was used to estimate the gas mass. The blue lines represent theoretical predictions by \cite{Vijayan19}.}
\vspace{-0.3cm}
\label{fig5}
\end{figure*}

Figure \ref{fig4} presents the dust-to-gas (DtoG) ratio versus stellar mass relation from redshift 0 to $\sim$ 5.5. Line styles and shaded areas have the same meaning as in figure \ref{fig3}, while symbols represent observational data from \cite{Grossi15}, \cite{Nersesian19}, and \cite{DeVis19}. Open symbols correspond to cases where only atomic hydrogen was used to estimate the total gas mass. Both DUSTY-GAEA and DUSTY-GAEA-Dwek models feature a weak increase in the DtoG ratio as a function of stellar mass. 

Nonetheless, they differ in their detailed trends. Both models reproduce the observational data in the local Universe well with a comparable scatter. In contrast, DUSTY-GAEA-NoG predicts a decreasing DtoG ratio with stellar mass except for redshift zero, where it significantly underpredicts the observed DtoG ratios.

Figure \ref{fig5} displays predictions for the dust-to-metal (DtoM = $\frac{M_d}{M_d + M_m}$, where M$_m$ is the total mass of metals in the gas phase) ratio normalised to the Milky Way value (i.e. 0.44) versus stellar mass. Line styles and shaded areas have the same meaning as  in figure \ref{fig3}, and symbols correspond to observational data from \cite{Grossi15} and \cite{DeVis19}. Open symbols indicate cases where only atomic hydrogen was used to estimate the gas mass. The blue lines represent model predictions by \cite{Vijayan19}, who included dust formation and evolution models in the semi-analytical galaxy formation model L-Galaxies. This model was run on the same N-body simulations considered in this study. The authors have studied in detail the DtoM ratio.

Following \cite{Popping17}, we estimated the total metal abundance using the 12+$log(\text{O/H})$ values reported by \cite{Grossi15} and \cite{DeVis19}. The DtoM ratio trends are slightly shallower than the DtoG ratio trends across all redshifts. Unlike the DtoG ratio, which decreases slightly with redshift in the fiducial model, the DtoM ratio increases slightly with redshift in both DUSTY-GAEA and DUSTY-GAEA-Dwek models. However, the DtoM ratio and the DtoG ratio share the same features. The DUSTY-GAEA and DUSTY-GAEA-Dwek models perform similarly well in comparison to the observational data. The model by Vijayan et al. predicts slightly shallower trends than our fiducial model and the opposite redshift evolution, i.e. the DtoM ratio trend decreases with redshift. Our model predicts a slightly higher DtoM ratio down to redshift $\sim$ 3.5, a similar ratio at $\sim$ 2.5 and 1.6, and a lower ratio in the local Universe. Generally, our model predicts much weaker redshift evolution.

\begin{figure*}
\centering
    \makebox[\textwidth][c]{\includegraphics[width=1.1\linewidth]{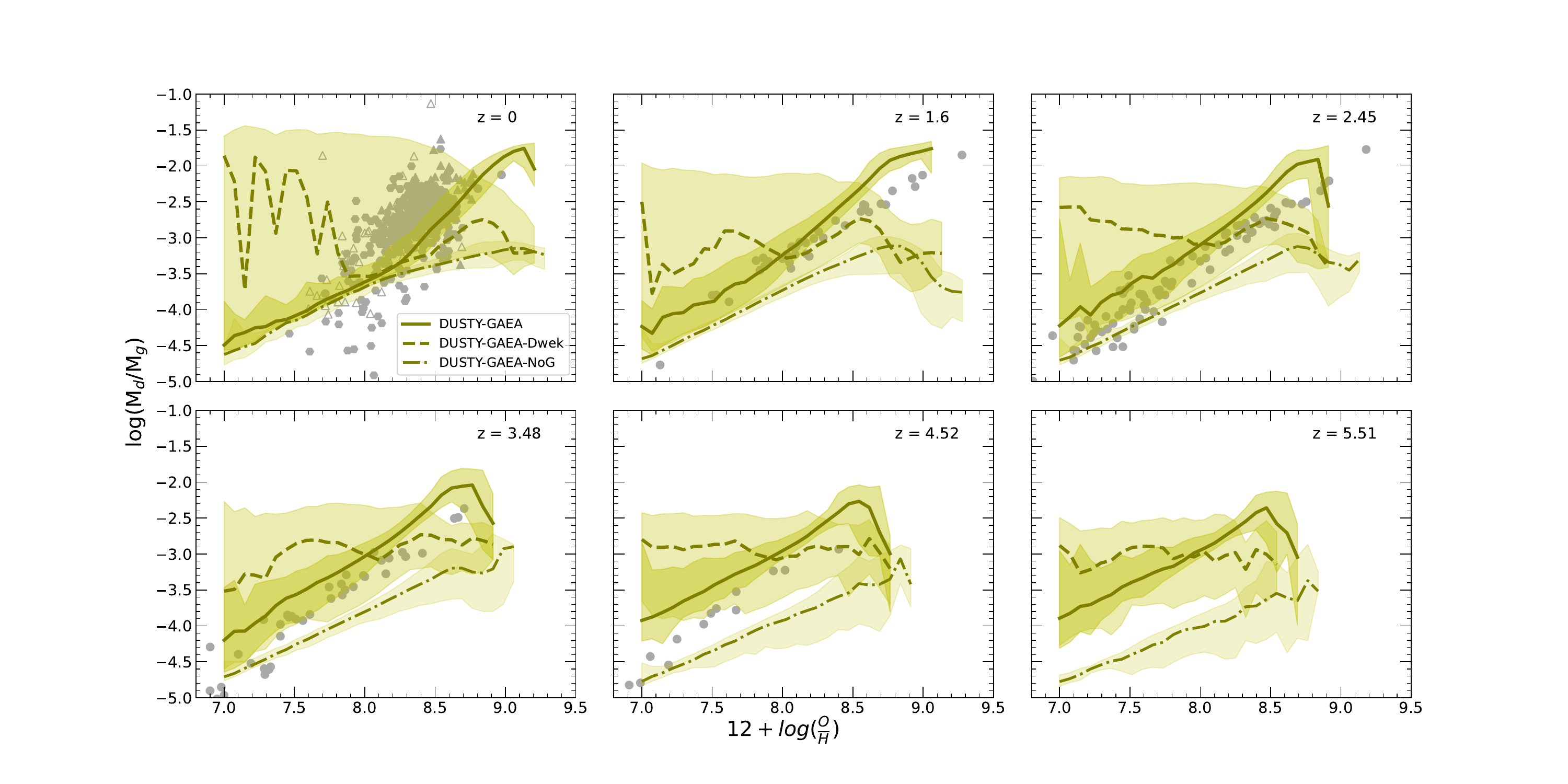}\par}      
    \caption{The dust-to-gas (DtoG) ratio as a function of metallicity at different redshifts. By metallicity, we refer to the gas-phase oxygen abundance. Line styles and shaded areas have the same meaning as in figure \ref{fig3}. Symbols represent the observational data by \cite{Grossi15} (up triangles) and \cite{DeVis19} (hexagons) at z $\sim$ 0, and \cite{Peroux20} (circles) at higher redshifts. Open symbols correspond to cases where only atomic hydrogen was used to estimate the total gas mass.}
\vspace{-0.3cm}
\label{fig6}
\end{figure*}

\begin{figure*}
\centering
    \makebox[\textwidth][c]{\includegraphics[width=1.1\linewidth]{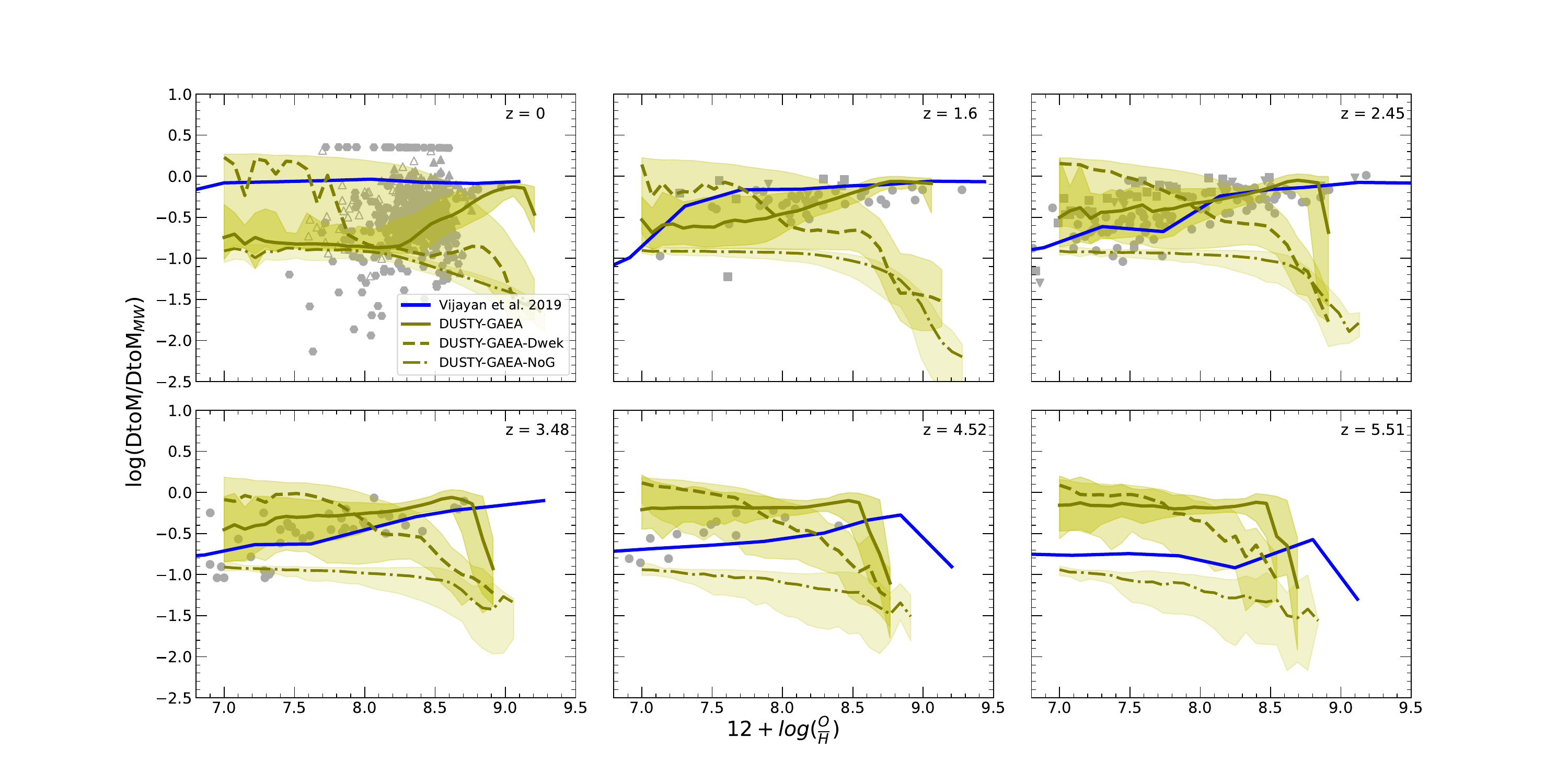}\par}      
    \caption{The dust-to-metal (DtoM) ratio normalized to the Milky Way value (i.e. 0.44) as a function of metallicity at different redshifts. Line styles and shaded areas have the same meaning as  in figure \ref{fig3}. Symbols correspond to observational data from \cite{Grossi15} (up triangles) and \cite{DeVis19} (hexagons) at z $\sim$ 0, and \cite{DeCia16} (squares), \cite{Wiseman17} (down triangles), and \cite{Peroux20} (circles) at higher redshifts. Open symbols indicate cases where only atomic hydrogen was used to estimate the gas mass. The blue lines represent theoretical predictions by \cite{Vijayan19}.}
\vspace{-0.3cm}
\label{fig7}
\end{figure*}

\subsection{Dust properties versus Metallicity}

Figure \ref{fig6} shows predictions for the DtoG ratio versus metallicity. By metallicity, we refer to the gas-phase oxygen abundance in units of 12 + log($\frac{O}{H}$). Line styles and shaded areas have the same meaning as  in figure \ref{fig3}, and symbols correspond to observational data from \cite{Grossi15} and \cite{DeVis19} at z $\sim$ 0, and \cite{Peroux20} at higher redshifts. Open symbols indicate cases where only atomic hydrogen was used to estimate the gas mass. The DtoG ratio behaves as expected, increasing with metallicity across all redshifts, in the models DUSTY-GAEA and DUSTY-GAEA-NoG. DUSTY-GAEA predicts a linear correlation down to redshift $\sim$ 3.5. Below redshift $\sim$ 3.5, the correlation becomes gradually steeper at high metallicity. Meanwhile, DUSTY-GAEA-NoG maintains an almost linear correlation across redshifts (a slope between 0.95 and 0.75, depending on the redshift). The bends seen in the high metallicity bins are due to destruction overtaking formation by stars, same as the bends seen in the DUSTY-GAEA predictions. 

 Contrary to DUSTY-GAEA and DUSTY-GAEA-NoG, the model DUSTY-GAEA-Dwek, shows almost no correlation between the DtoG ratio and metallicity and a significantly large scatter. This is mainly because in this model, oxygen abundance is no longer a good indicator of galaxy metallicity. The Dwek model only specifies the dust-forming elements (e.g. O, Fe, Mg) and not in what species of dust grains these elements reside (e.g. Olivine and Pyroxene). Therefore, the amount of oxygen allowed to be incorporated in dust grains is not constrained, resulting in over-depletion of oxygen. If we use the mass of all the metals on the x-axis instead, the DtoG ratio would increase with metallicty as expected.

The behaviour of the DUSTY-GAEA and DUST-GAEA-Dwek models on the plane of the DtoG ratio versus metallicity is the main reason we choose DUSTY-GAEA as our fiducial model. Furthermore, DUSTY-GAEA predictions of the oxygen depletion fraction ($\frac{M_{o,d}}{(M_{o,d} + M_{o,g})}$ $\sim$ 0.03 - 0.2) are consistent with those adopted in emission-line modelling (e.g. \citealp{Groves04}; \citealp{Gutkin16}), unlike predictions from DUSTY-GAEA-Dwek ($\sim$ 0.03 - 0.95). When comparing our model predictions with observations, it is worth noting that the observational data presented in this figure (and the next) includes measurements using emission lines in the local Universe and absorption lines at high redshifts. Besides the systematic difference between these measurements, one should also keep in mind that it is not well understood which galaxies give rise to DLAs at high redshifts. Hence, the comparison with our model predictions at high redshifts should be interpreted with caution.
\begin{figure*}
\centering
     \makebox[\textwidth][c]{\includegraphics[width=1.1\linewidth]{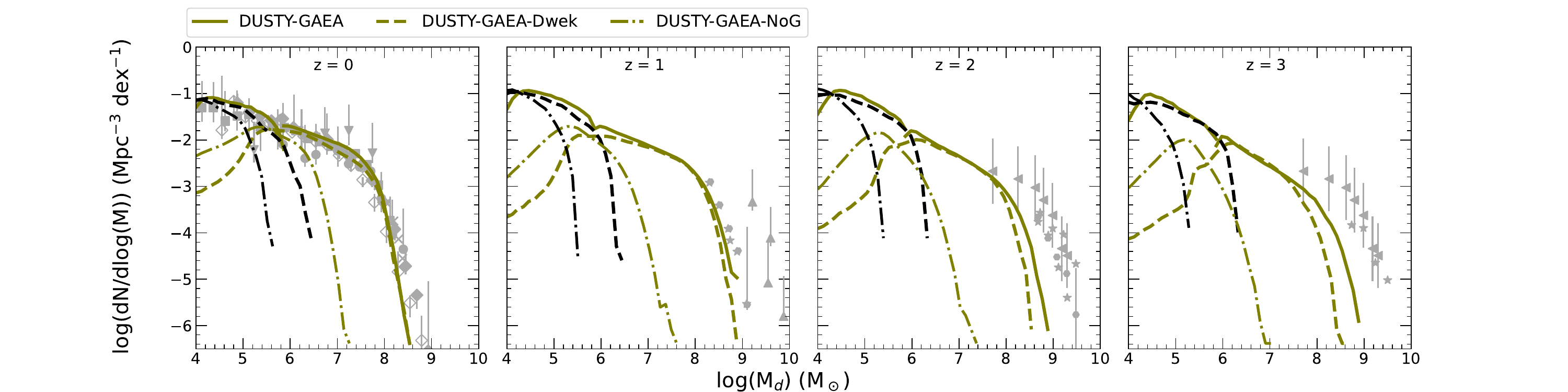}\par}      
    \caption{The dust mass function (DMF) from redshift $\sim$ 0 to 3. Solid, dashed, and dashed-dotted lines represent predictions by DUSTY, DUSTY-GAEA-Dwek, and no dust growth models, respectively. Olive and black lines represent predictions based on the MSI and MSII merger trees, respectively. Symbols represent a compilation of observational data from \cite{Dunne03} (left triangles), \cite{Vlahakis05} (diamonds), \cite{Eales09} (up triangles), \cite{Clemens13} (crosses), \cite{Beeston18} (squares), \cite{Clark15} (down triangles), \cite{Dunne11} (circles), \cite{Pozzi20} (hexagons), and \cite{Traina24} (stars).}
\vspace{-0.3cm}
\label{fig8}
\end{figure*}

In the local Universe, predictions of our fiducial model (DUSTY-GAEA) underestimate the DtoG ratio of the bulk of the observational data by $<$ 0.5 dex. However, looking at figure \ref{fig4},  the DtoG ratio is well reproduced as a function of stellar mass, making us wonder whether this shortcoming is due to metallicity overestimation instead (see arguments by \cite{Fontanot21} and \cite{DeLucia24} regarding the mass-metallicity relation predictions from GAEA). Indeed, the transition metallicty to the steep slope of the correlation in DUSTY-GAEA is 8.3 dex, about 0.2 dex higher than the one measured by \cite{Remy14}. Shifting the model by 0.2 dex leftward would align the model well with the data beyond a metallicity of 8 dex. Below 8 dex, the model slightly overestimates the DtoG ratio. One could argue that overestimation of the metallicity (indicated by the oxygen abundance) should reflect on the DtoG ratio since it increases as a function of (the total) metallicity. However, one should remember that only a small fraction of oxygen is incorporated into dust grains (low depletion $<$ 0.2). Moreover, the 0.2 dex shift could easily be accounted for by the large systematic uncertainties on the metallicity estimates ($\sim$ 0.7 dex, \citealp{Kewley08}; \citealp{Hirschmann23}). At high redshift, the model slightly overestimates the DtoG ratio; however, it predicts the behaviour reasonably well.

Figure \ref{fig7} presents the DtoM ratio normalised to the Milky Way value of 0.44 versus metallicity. Line styles and shaded areas have the same meaning as  in figure \ref{fig3}, and symbols correspond to observational data from \cite{Grossi15} and \cite{DeVis19} at z $\sim$ 0, and \cite{DeCia16}, \cite{Wiseman17}, and \cite{Peroux20} at higher redshifts. Open symbols indicate cases where only atomic hydrogen was used to estimate the gas mass. The blue lines represent model predictions from \cite{Vijayan19}. The DtoM ratio in DUSTY-GAEA and DUSTY-GAEA-NoG maintain a rather weak/no correlation with metallicty down to redshift $\sim$ 3.5. Below redshift $\sim$ 3.5, the DUSTY-GAEA model predicts a clear correlation with metallicity at the high-metallicity end, in good agreement with the available constraints. On the other hand, the DUSTY-GAEA-Dwek and DUSTY-GAEA-NoG predict an anti-correlation with metallicity, at variance with the data. 

The same arguments made for the DtoG could be made here for the comparison with the observational data. However, DUSTY-GAEA reproduces the data at high-z better on this plane. In contrast to our models,  Vijayan et al. models predict a correlation between the DtoM ratio and metallicity at low metallicity and no correlation at high metallicity. This kind of behaviour changes with redshift in such a way that the no correlation segment extends down to lower metallicities as redshift decreases. Accordingly, predictions of this model overestimate the data across the metallicity range in the local Universe, reproduce the high metallicity end at redshift $\sim$ 1.6, are broadly consistent with the data at redshifts $\sim$ 2.45 and 3.48, and reproduce the low metallicity end at redshift $\sim$ 4.52.

\begin{figure}
\centering
    \includegraphics[height=8.cm,width=8.cm]{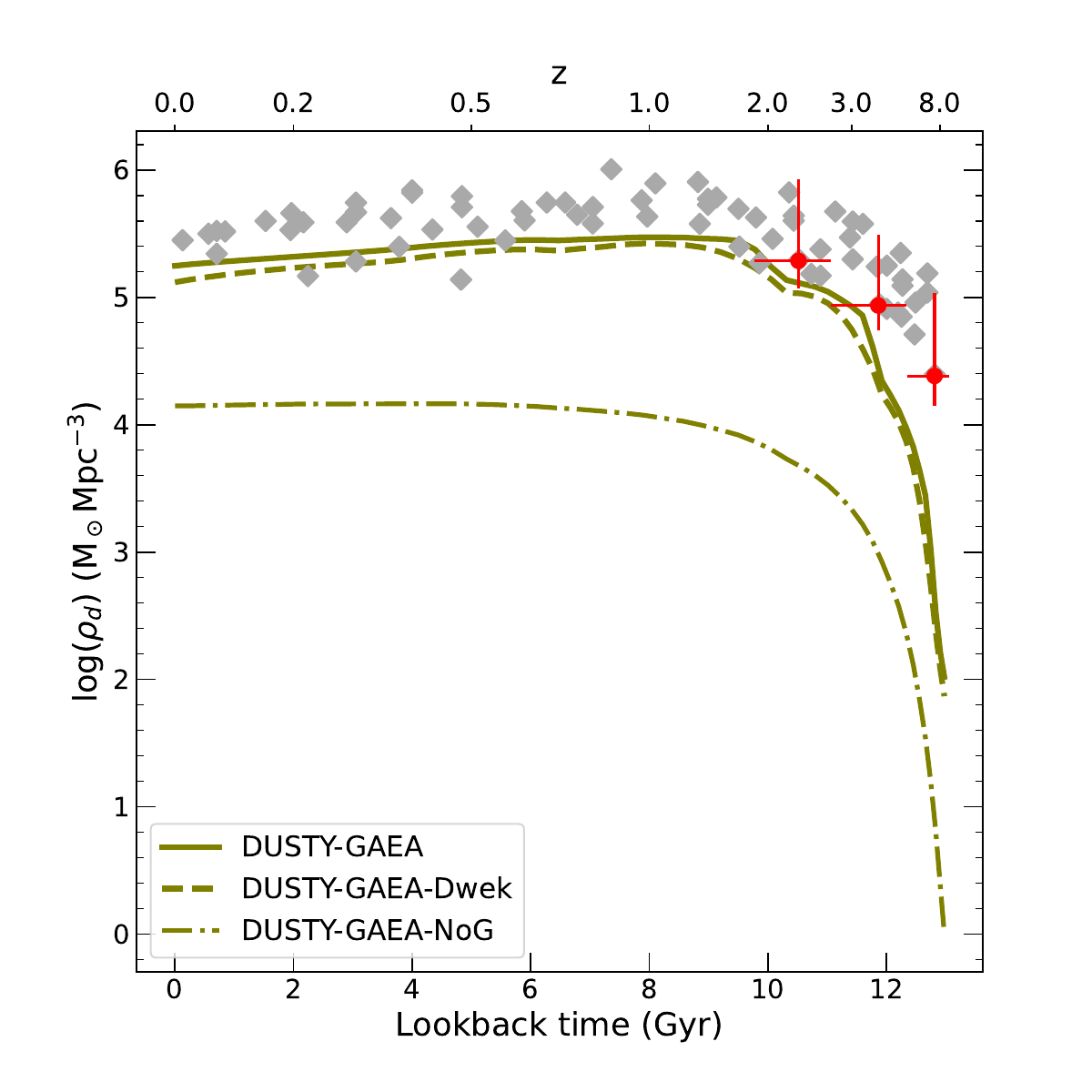}\par      
    \caption{Predictions of the cosmic density of dust (CDD) as a function of the lookback time (bottom x-axis) and redshift (top x-axis). Solid, dashed, and dashed-dotted lines represent predictions by DUSTY-GAEA, DUSTY-GAEA-Dwek, and DUSTY-GAEA-NoG models, respectively. Gray diamonds represent data collected by \cite{Berta25}, while the red dots represent their measurements.}
\vspace{-0.3cm}
\label{fig9}
\end{figure}

\begin{figure*}
\centering
    \makebox[\textwidth][c]{\includegraphics[width=1.1\linewidth]{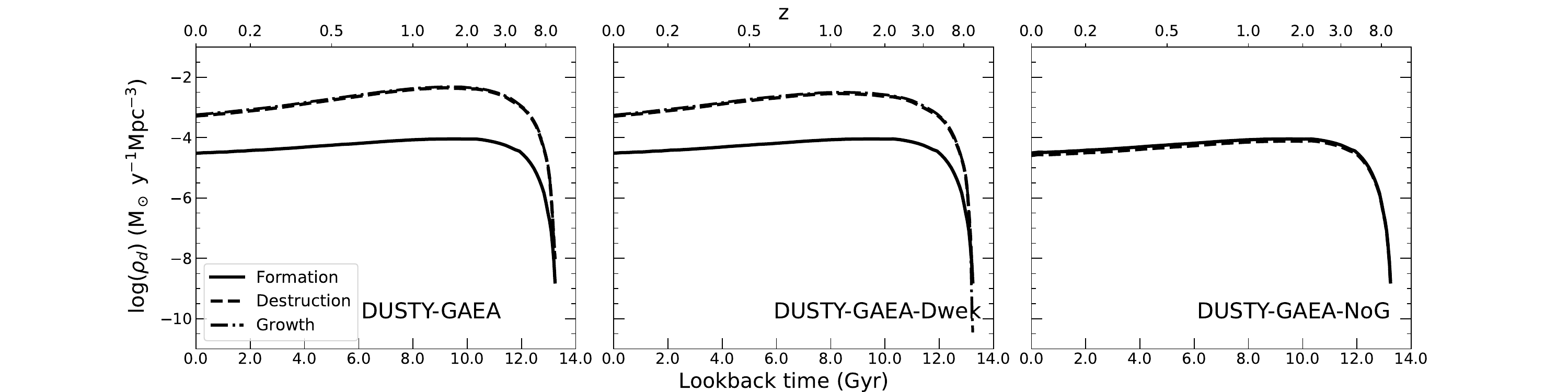}\par}      
    \caption{Predictions of the cosmic density of dust formation and destruction rates. Line styles have the same meaning as  in figure \ref{rates}.}
\vspace{-0.3cm}
\label{rates-ev}
\end{figure*}

\subsection{The dust mass function and cosmic evolution}

DUSTY-GAEA (solid), DUSTY-GAEA-Dwek (dashed), and DUSTY-GAEA-NoG (dashed-dotted) model predictions of the dust mass function (DMF) from redshift $\sim$ 0 to 3 are shown in figure \ref{fig8}. Olive and black lines represent predictions based on the MSI and MSII merger trees, respectively. Symbols represent a compilation of observational data from \cite{Dunne03}, \cite{Vlahakis05}, \cite{Eales09}, \cite{Clemens13}, \cite{Beeston18}, \cite{Clark15}, \cite{Dunne11}, \cite{Pozzi20}, and \cite{Traina24}.  We remind the reader that errors and uncertainties on the dust mass estimates discussed in Section \ref{dust-mass} would propagate into the DMF and should be kept in mind while comparing our model predictions to observations.

At z $\sim$ 0, DUSTY-GAEA and DUSTY-GAEA-Dwek reproduce the low mass end of the DMF and the position of the knee rather well; however, they slightly underestimate the high mass end. The fiducial model has a marginally higher number density than the DUSTY-GAEA-Dwek variant at intermediate dust masses. The two models behave similarly at z $\sim$ 1, and they are only consistent with \cite{Pozzi20} and \cite{Traina24} observations. They underestimate the observations by  \cite{Eales09}.  \cite{Eales09} observations are based on data obtained by the Balloon-borne Large Aperture Submillimeter Telescope (BLAST), which could be limited by the small sample size and difficulties of ground observations. Above reshift one, DUSTY-GAEA has a considerably higher number density above dust mass of 10$^{7.5}$ M$_\odot$ compared to the DUSTY-GAEA-Dwek variant, but still predicts number density below the observational constraints. Contrary to DUSTY-GAEA and DUSTY-GAEA-Dwek models, DUSTY-GAEA-NoG significantly underestimates the DMF at all redshifts, except for the low mass end at redshift zero. 

Both DUSTY-GAEA and DUSTY-GAEA-Dwek predict very mild redshift evolution of the DMF, where the DMF increases between redshifts three and one and decreases between one and zero, consistent with observations (see, e.g. \citealp{Beeston24}). DUSTY-GAEA-NoG predicts an increase in the dust mass function between redshift three and zero, regardless of the galaxy dust mass, similar to the evolution predicted by the fiducial model of \cite{Popping17}.

Figure \ref{fig9} shows DUSTY-GAEA (solid), DUSTY-GAEA-Dwek (dashed), and DUSTY-GAEA-NoG (dashed-dotted) predictions of the cosmic density of dust (CDD). Gray diamonds represent data collected by \cite{Berta25}, while the red dots represent their measurements. The 1$\sigma$ scatter of Berta et al. measurements represents kind of an upper limit of the average scatter on the y-axis (about 0.5 dex). Our models,  DUSTY-GAEA and DUSTY-GAEA-Dwek, reproduce well the shape of the CDD with a normalization slightly lower than the data but consistent with the 1$\sigma$ scatter. This underestimation is expected since our models are not able to reproduce well the dust abundance in the most massive galaxies at high redshift. The model DUSTY-GAEA-NoG neither reproduces well the bend of the CDD below redshift one nor the normalization.

Figure \ref{rates-ev} shows our model predictions of the cosmic density of dust formation and destruction rates. Line styles have the same meaning as  in figure \ref{rates}. All rates, formation by stars, destruction, and growth, increase with decreasing redshift to reach a maximum between z $\sim$ 2 and 1  before decreasing towards redshift zero, consistent with the behaviour of the cosmic density of star formation rate (\citealp{Madau14}). In the redshift range we present here, dust formation via growth in the ISM dominates the dust cosmic density, having about 1.2 and 1.6 dex higher rates in DUSTY-GAEA and DUSTY-GAEA-Dwek, respectively, compared to the rates estimated for the formation by stars. However, in individual galaxies, growth might not always be strongly dominant (see figure \ref{rates} and figure \ref{ratesA}).

%%%%%%%%%%%%%%%%%%%%%%%%%%%%%%%%%%%%%%%%%%%%%%%%%%%%%%%%%%%%%%%%%%%%%%%%%%%%%%%%%%%%%%%%%%%%%%%%%%%%%%%%%%%%%%%%%%%%%%%%%%%%%%%%%%%%%%%%%%%%%%%%%%%%%%%%%%%%%%%%%%%%%%%%%

\section{Discussion}
\label{Discussion}

Theoretical studies of dust formation and evolution, coupled with galaxy evolution, are highly needed to decode the large amount of information contained in the observed dust scaling relations (e.g. \citealp{Inoue03}; \citealp{Zhukovska08}; \citealp{Remy14}; \citealp{Algera25}). For instance, \cite{Inoue03} showed that for local galaxies, the DtoG ratio versus metallicity relation does not represent an evolutionary sequence where galaxies increase their dust and metal content over time, keeping the DtoM ratio constant, but rather a sequence in which galaxies have similar ages but different star formation histories. \cite{Remy14} also argued that the scatter in this relationship encodes information about the galaxy star formation histories, dust destruction efficiency, grain size distribution, and chemical composition. Besides interpreting observations, modelling galaxy physical properties (e.g. colours, H$_2$ and HI fractions) requires some degree of dust physics treatment.  

The field of dust physical modelling has matured over the past few decades, with dust formation and evolution processes implemented in several one-zone models (e.g. \citealp{Dwek80}; \citealp{Mckee89}; \citealp{Dwek98}; \citealp{Inoue03}; \citealp{Zhukovska08}; \citealp{Hirashita15}), semi-analytic models (e.g. \citealp{Popping17}; \citealp{Vijayan19}; \citealp{Triani20}; \citealp{Dayal22}; \citealp{Parente23}; \citealp{Yates24}), and hydrodynamic simulations (e.g. \citealp{Bekki13}; \citealp{McKinnon16}, \citealp{McKinnon17}; \citealp{Aoyama17}, \cite{Aoyama18}; \citealp{Hou17}, \citealp{Hou19};  \citealp{Gjergo18}; \citealp{Li19}; \citealp{Granato21}). All of these frameworks agree on the central role played by dust growth in the ISM. In the previous section, we demonstrated that and showed that our fiducial model reproduces multiple observational constraints. Our model also predicts dust growth to be the main dust formation mechanism in the Universe up to z $\sim$ 8. This is broadly consistent with results from \cite{Popping17}, \cite{Vijayan19}, and \cite{Yates24}, but in contrast with predictions from \cite{Triani20} who find growth to be dominat only up to z $\sim$ 1.

\subsection{Comparison with other semi-analytical models}
\label{Compa_M}
In this section, we compare our fiducial model predictions to those from the semi-analytical models by \cite{Popping17}, \cite{Vijayan19}, \cite{Triani20}, \cite{Parente23}, and \cite{Yates24}. Note that predictions from these models are made imposing selection criteria different from ours, except for the predictions by Triani et al.\footnote{We limit ourselves to a comparison with semi-analytical models because of the use of the same theoretical framework makes the comparison more straightforward. We compare our DMF predictions with predictions from hydrodynamical simulations in Appendix \ref{appC}.}

All these models include explicit treatments for dust formation by stars, growth in the dense ISM, destruction by SNe forward shocks, and sputtering by the hot gas. Additionally, they account for the dust locked into stars when formed (i.e. astration) as well as dust recycling as part of the baryon cycle. Furthermore, all these models are run on the same (MSI) merger trees, except for the model by Popping et al., that is run on merger trees obtained using the extended Press-Schechter (EPS). Vijayan et al. and Yates et al. also provided predictions from the model run on the MSII merger trees. The specific implementations of the processes governing dust evolution vary as briefly summarized in Table \ref{table3}.

\begin{table*}
\centering
\caption{Summary of the dust physics prescriptions adopted in the semi-analytical models considered in this work for comparison with our predictions.}
\begin{tabular}{lllllll}
\hline
{Semi-analytical model}
& {Dust formation}
& {Dust destruction}
& {Dust growth$^g$}
& {Dust sputtering} \\
\hline
{Santa Cruz} & SNIa, SNII, AGBs & SNe forward shocks$^c$ & Model by& In the hot gas \\
{\cite{Popping17}} & {\cite{Dwek98}}$^a$ & {\cite{Mckee89}}& {\cite{Zhukovska08}}& {\cite{Tsai95}} \\
\hline
{L-Galaxies} & SNII, AGBs & SNe forward shocks$^c$ & Model by& In the hot gas \\
{\cite{Vijayan19}} & {\cite{Zhukovska08}}$^a$  & {\cite{Mckee89}}& {\cite{Zhukovska08}}& Completely destroyed\\
 &  {\cite{Ferrarotti06}}$^b$& & & \\
\hline
{DUSTY-SAGE} & SNII, AGBs & SNe forward shocks$^d$ & Model by& In the hot gas \\
{\cite{Triani20}} & {\cite{Dwek98}}$^a$  & {\cite{Mckee89}}& {\cite{Dwek98}}& {\cite{Tsai95}}\\
\hline
{L-Galaxies} & SNII, AGBs & SNe forward shocks$^d$ & Model by& In the hot gas \\
{\cite{Parente23}} & {\cite{Dwek98}}$^e$  & {\cite{Mckee89}}& {\cite{Hirashita11}}& {\cite{Tsai95}}\\
\hline
{L-Galaxies} & SNIa, SNII, AGBs & SNe forward shocks$^c$ & Model by& In the hot gas \\
{\cite{Yates24}} & {\cite{Zhukovska08}}$^a$  & {\cite{Mckee89}}& {\cite{Zhukovska08}}& {\cite{Tsai95}}\\
 &  {\cite{Ferrarotti06}}$^b$& & & \\
\hline
{DUSTY-GAEA} & SNIa, SNII, AGBs & SNe forward shocks$^f$ & Model by& In the hot gas \\
{This work} & {\cite{Dwek98}}$^a$ & {\cite{Mckee89}}& {\cite{Zhukovska08}}& {\cite{Tsai95}} \\
\hline
\end{tabular}\\
$^a$ Formulae.
$^b$ Dust yield tables.
$^c$Assuming a fixed mass of gas cleared of dust per SN event.
$^d$Assuming a swept up mass by the shock depending on the metallicity of the ambient gas \cite{Yamasawa11}.
$^e$Combined \cite{Dwek98} equations with the key element concept from \cite{Zhukovska08} to preserve Olivine chemical composition.
$^f$ Assuming a swept up mass by the shock depending on the density of the ambient gas.
$^g$Although all these models, except DUSTY-SAGE, use the same growth model or employ the same key element concept (\citealp{Hirashita11}), the details of specific implementations are different.

\label{table3}
\end{table*}

\begin{figure*}
\centering
   \makebox[\textwidth][c]{\includegraphics[width=1.1\linewidth]{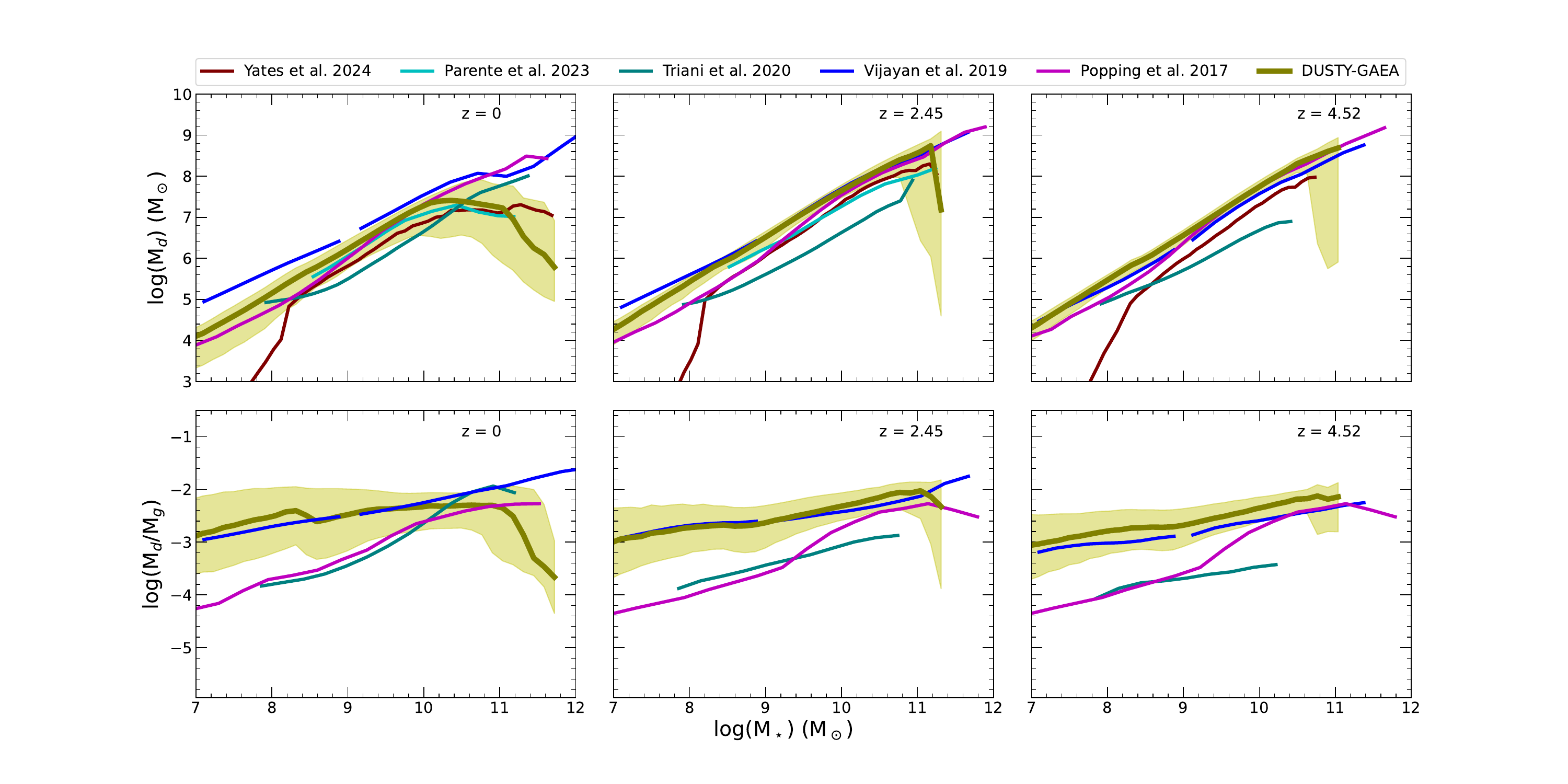}\par}    
    \makebox[\textwidth][c]{\includegraphics[width=1.1\linewidth]{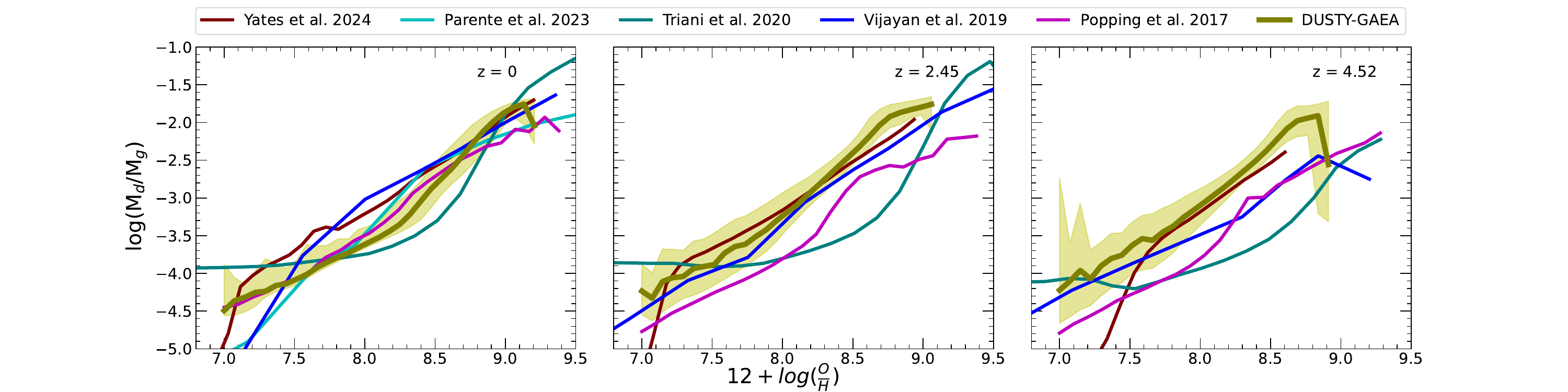}\par}      
    \caption{The dust mass (top row) and the DtoG ratio (second row) as a function of the stellar mass from redshift $\sim$ 0 to 4.5, and the DtoG ratio (bottom row) as a function of metallicity at the same redshifts. Coloured solid lines represent the predictions of the semi-analytical models by \cite{Popping17} (magenta), \cite{Vijayan19} (blue), \cite{Triani20} (teal), \cite{Parente23} (cyan), and \cite{Yates24} (maroon).}
\vspace{-0.3cm}
\label{fig10}
\end{figure*}

\underline{Scaling relations}: Figure \ref{fig10} shows the dust mass (top row) and the DtoG ratio (second row) as a function of the stellar mass from redshift $\sim$ 0 to 4.5, and the DtoG ratio (bottom row) as a function of metallicity at the same redshifts. Coloured solid lines show predictions from different models, as indicated in the legend. Our model predictions of the dust mass in galaxies are broadly consistent with all models from redshifts $\sim$ 0 to 4.5, except for the model by Triani et al., whose dust mass falls notably below the predictions of the other models as the redshift increases. This is due to the fact that, unlike the other models, Triani et al. find that dust growth in the dense ISM is the dominant dust formation mechanism only below redshift $\sim$ 1. At the low stellar mass end ($<$ 10$^8$ M$_\odot$), predictions by Yates et al. are significantly below all models. This is mainly because of their dust growth formalism, that includes a timescale dependent on the dust abundance, leading to an extremely inefficient dust growth at low stellar masses. In this range, dust is basically only formed via SNII (R. Yates, private communication), which is also reflected in the DtoG ratio versus metallicity relation (see below).

It is interesting how the broad consistency between the models seen in the dust mass-stellar mass relation breaks down when relations such as the DtoG versus stellar mass and metallicity are considered. Our predictions of the DtoG ratio as a function of the stellar mass are consistent only with  predictions by Vijayan et al. at all redshifts. The DtoG rations predicted by Popping et al. are lower than ours, except at the high stellar mass end ($\sim$ $10^{10}$ M$_\odot$), indicating larger gas reservoirs in their simulated galaxies compared to ours. Predictions by Triani et al. are consistent with our predictions only at the high stellar mass end in the local Universe.

The DtoG ratio versus metallicity is an excellent tracer of dust processing in the ISM, and represents a strong constraint on models of dust formation and evolution (e.g. \citealp{Hirashita99}; \citealp{Kuo12}; \citealp{Hirashita13}; \citealp{Asano13a}; \citealp{Remy14}). In this plane, each model considered behaves more or less differently and they are all broadly consistent only at redshift zero, where all the models are somewhat calibrated to reproduce some observational constraints (e.g. the galaxy stellar mass function). At higher redshift, our predictions remain consistent only with those by Vijayan et al. and Yates et al. at z $\sim$ 2.5 and Yates et al. at metallicities larger than $\sim$ 7.5 at z $\sim$ 4.5.  

These models have different critical metallicities at which dust growth becomes the dominant dust formation mechanism. This metallicity is set by the efficiency of dust growth and the star formation history (see, e.g. \citealp{Inoue03}; \citealp{Asano13a}). \cite{Popping22} suggested that these differences are driven by differences in the adopted time scale of star formation. We argue instead that these are driven, at least in part,  by differences in the specific dust growth models adopted. For instance, the model adopted by Yates et al. strongly suppresses dust growth in metal-poor low-stellar mass galaxies compared to the model by Popping et al. (see their figures for growth timescales). This results in a sharp decrease in the DtoG ratio at low metallicity in the former model, marking the critical metallicity. The long growth timescale in Triani et al.  likely pushes their critical metallicity towards higher values. 

\begin{figure*}
\centering
     \makebox[\textwidth][c]{\includegraphics[width=1.1\linewidth]{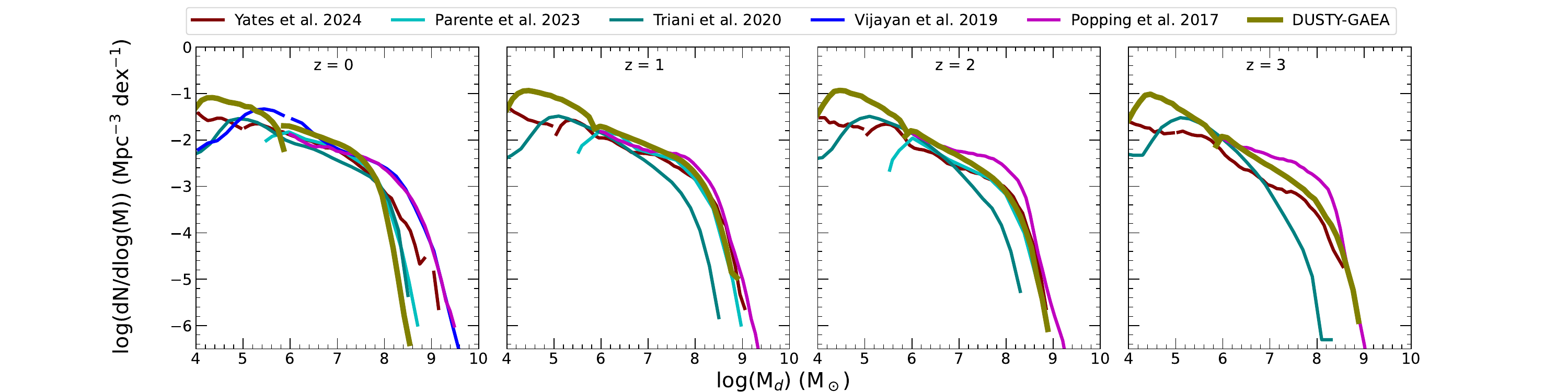}\par}     
    \caption{The dust mass function (DMF) from redshift $\sim$ 0 to 3. Solid olive lines represent predictions by DUSTY-GAEA, combining the MSI and MSII runs. Coloured solid lines represent predictions from the semi-analytical models described in figure \ref{fig10}.}
\vspace{-0.3cm}
\label{DMF-m}
\end{figure*}

\underline{Cosmic dust evolution}: Figure\ref{DMF-m} shows predictions from our reference model of the dust mass function (DMF) from redshift $\sim$ 0 to 3, and compares them to predictions from other semi-analytic models discussed. In the local Universe, our model predictions are consistent with predictions by Triani et al., Parente et al., and Yates et al. above a dust mass of 10$^6$ M$_\odot$. The models by Popping et al. and Vijayan et al. predict a knee at larger dust masses. At $z\sim 1$ and dust mass above 10$^6$ M$_\odot$, all models are consistent with each other, except the model by Triani et al. that predicts lower number densities of galaxies at fixed dust mass. At higher redshift, the number densities of galaxies around the knee decrease significantly in all models, except for the model by Popping et al. 

At large dust masses, differences between the models are driven both by the specific implementation of dust physics and a different simulated volume. As mentioned above, all models are based on the same dark matter simulation but Popping et al. that uses analytic merger trees built using the extended PS formalism. This might explain, at least in part, the larger number densities of dust-rich galaxies found in this model at high redshift. As for the differences visible at low dust masses, we note that the models by Triani et al. and Parente et al. have only been run over the MSI merger trees while the models by Vijayan et al. and Yates et al. can resolve galaxies down to a stellar mass of  $\sim$ 10$^7$ M$_\odot$ as they have also been run on the higher resolution MSII simulation.

\underline{Dust physics}: Our model without dust growth underpredicts the scaling relations at all redshifts (see Section \ref{results}), and the trends of some of the relations are inconsistent with the observed ones. Predictions of such a model could be improved if a larger dust condensation efficiency in the stellar ejecta is assumed, as demonstrated by \cite{Popping17} (their model without dust growth). However, the authors also showed that such implementation results in a significant overestimation of the DtoG and DtoM ratios as a function of metallicty in metal poor galaxies. In the framework of our model, a high condensation efficiency is not compatible with the stellar yields we adopt because the amount of oxygen ejected is not enough to sustain such condensation efficiencies (\citealp{McKinnon16} reported a similar problem). 

Several studies have discussed possible variations of the dust condensation efficiency with e.g. the progenitor mass and metallicity (e.g. \citealp{Ferrarotti06}; \citealp{Zhukovska08}; \citealp{Valiante09}; \citealp{Ventura12a}; \citealp{Gioannini17}). However, the influence of such variations on the overall dust scaling relations is still not fully understood. For instance, \cite{Calura08}, using one-zone chemical evolution models, showed that variations in the condensation efficiency in SNeII ejecta  (from 0.1 to 1) have an insignificant effect when dust growth is included. The same could be deduced from the results by \cite{Popping17} at large stellar masses (comparing their fiducial model and the model with high condensation). At the low stellar mass end, changing the condensation efficiency has a strong impact on model predictions. This is probably because the higher the stellar mass is, the more dust growth dominates dust formation, and dust grains are less likely to have memory of their stellar origins (see their figure B3). Changing the stellar metal yields and the assumed IMF will also change the amount of dust formed by stars (we will address these issues in future work). 

Variations in the condensation efficiency could also have an important impact on galaxies at high redshifts because at early epochs, dust formation by stars is expected to play an important role. Current observational constraints are, however, sparse and include extreme objects (e.g. extremely star-forming or AGN hosts). These objects might not be the best candidates to constrain dust formation models since they might have had an accelerated evolution, making their dust contents contaminated by growth (e.g. \citealp{Algera25}; \citealp{Faisst25}; \citealp{Nanni25}; \citealp{Osman25}). 

One could also argue that at high redshift dust destruction becomes less efficient. This could be the case because of the larger gas densities, which prevents supernovae remnants from propagating far into the ISM (e.g. \citealp{Nozawa06}; \citealp{Hirashita13}). However, we find that in the framework of our model assuming destruction timescales dependent on gas density or metallicity leads to results very similar to those of our  fiducial model in massive galaxies ($>$ 10$^{8.5}$ M$_\odot$, see Appendix \ref{appB}). One aspect that we have not considered, and that might be relevant, is that there could be variations in the fraction of dust destroyed by a single SNa (e.g. \citealp{Nozawa06}; \citealp{Zhukovska08}).

\subsection{Limitations of our model}
\label{limitations}
Estimating the gas density where dust growth and destruction occur is challenging, but essential. \cite{Popping17} already pointed out the necessity of having a varying growth timescale as a function of the gas density for reproducing the cosmic dust evolution (i.e. simultaneously reproducing the dust mass of low- and high-redshift galaxies). This is because a short growth timescale is needed at high redshifts, which could not be achieved depending only on the (low)metallicity during those epochs. We followed their lead, but adopted a different approach for estimating the gas density. Our approach uses the estimated H$_2$ mass (see \citealp{Xie17} for the HI-H$_2$ partitioning in our model) and a fixed value of H$_2$ volume-filling factor to estimate the volume occupied by the molecular gas (\citealp{Draine11}; see Section \ref{SNeDes} for further details). The volume-filling factor is likely not the same in all galaxies at all redshifts; hence, a varying volume-filling factors would be more realistic.

Another limitation of our dust physical modelling is the lack of an explicit treatment for the grain size distribution, and associated processes of shattering and coagulation. Stellar sources mainly produce large grains ($\sim 0.1$ $\mu$m), and once they are injected into the ISM, the evolution of their size distribution is shaped by destruction, growth, shattering and coagulation (\citealp{Hirashita13}). Small grains due to their large surface area favour destruction and growth. Meanwhile, large grains are more likely to shatter into smaller grains or coagulate to form larger ones (e.g. \citealp{Hirashita13}; \citealp{Aoyama20}; \citealp{Parente23}). \cite{Hirashita12} argued that the effect of coagulation on the total dust mass budget could be neglected. Indeed, our model predictions are very similar to those by \citet{Parente23} who included a treatment of the grain size distribution based on the two-size approximation (\citealp{Hirashita15}). In future work, we plan to address this in the framework of our model, expanding it to include a treatment for grain size.

There are several channels through which dust could actively influence galaxy evolution (e.g. review by \citealp{Dorschner95}). For instance, molecular hydrogen forms most efficiently on the surfaces of dust grains  (e.g. \citealp{Hirashita02}; \citealp{Cazaux04}; \citealp{Wakelam17}), providing star formation with its necessary fuel (e.g. \citealp{Bigiel08}; \citealp{Fukui10}; \citealp{Schruba11}). Dust depletes a considerable amount of metals from the gas phase (e.g. \citealp{Zhukovska16}; \citealp{Zhukovska18}), affecting the ability of gas to cool down. Additionally, dust heats the ISM via processes such as the photoelectric heating (PEH), which is the primary heating process in the cold neutral and diffuse atomic hydrogen regions (\citealp{Wolfire95}; \citealp{Ingalls02}). PEH suppresses star formation, while enhancing SNe feedback (\citealp{Forbes16}; \citealp{Hu17}; \citealp{osman20b}). Dust also plays a role in enriching the circumgalactic medium through dusty outflows (\citealp{Ferrara91}; \citealp{Aguirre01}; \citealp{Bianchi05}; \citealp{Bekki15}; \citealp{Hirashita19}; \citealp{Kannan22}). In short, accounting for dust influence on galaxies involves accounting for all or some of these processes. In the current version of our model, we account only for metal depletion and the enrichment of the circumgalactic medium as part of the baryon cycle; inherently, this influences the gas cooling. Future developments of the model would include some of these other processes. In particular, molecular hydrogen formation on dust grains will have an impact on the star formation rates of our simulated galaxies since our star formation recipe depends on the molecular hydrogen abundance (\citealp{Xie17}).

\section{Summary and Conclusions}
\label{summary}

In this paper, we present a novel implementation of dust formation by stars, dust destruction by supernovae shocks and hot gas, and growth within the dense interstellar medium in the GAlaxy Evolution and Assembly theoretical model (GAEA). Our analysis demonstrates that the model successfully reproduces a wide range of observational constraints: the buildup of dust as a function of stellar mass out to $z\sim 6$, the scaling relations between the dust-to-gas/dust-to-metal ratio and stellar mass/metallicity in the local Universe, and the dust mass function out to $z\sim 1$.

A key outcome of our study is the dominant role of dust growth in the dense ISM, which emerges as the primary contribution to the cosmic dust budget to $z\sim 8$. Without such efficient growth, the observed dust abundances at high redshift cannot be explained. At the same time, the model highlights persistent difficulties in reproducing the number densities of dust-rich galaxies at $z>4$. These findings align with the broader theoretical landscape, suggesting that current prescriptions for stellar yields, condensation efficiencies, or ISM conditions remain incomplete or should be improved/modified at early cosmic epochs.

In this regime, the young age of the Universe challenges efficient dust formation via the growth mechanism, and dust formation by stars is expected to dominate the dust mass budget. However, current observational constraints mostly include massive galaxies for which growth might have played a role due to the earlier formation times. Observational measurements for low mass galaxies (M$_\star$ < 10$^9$ M$_\sun$) at high redshifts are necessary to improve our understanding of dust production at early times.

From a theoretical perspective, our study highlights the need for refined theoretical treatments that include grain size distribution, as well as metallicity dependant condensation efficiencies. A self-consistent modeling of molecular hydrogen formation on dust grains could also help to break model degeneracies and provide more robust predictions to be tested against the ever increasing amount of observational data we are gathering, particularly at early cosmic epochs.

\begin{acknowledgements}
We are also grateful to G. Popping, A. Vijayan, D. Triani, M. Parente, and R. Yates for making predictions of their models available to us. We are also grateful to M. Parente, R. Yates, H. Hirashita, M. Pannella, and S. Cristiani for stimulating discussions. We acknowledge support from the INAF 2022 Theory Grant ``The critical role of DUST in the cosmic barYon cycle”. OO and FF acknowledge support from the INAF 2024 RSN1 Minigrant ``The eﬀect of a variable IMF on Galaxy Evolution and Assembly: from the local to the high-z Universe”. This work has been partially supported by the Italian Ministry of University and Research (MUR) Missione 4 ``Istruzione e Ricerca” - Componente C2, Investimento 1.1 Fondo per il Programma Nazionale di Ricerca e Progetti di Rilevante Interesse Nazionale (PRIN), the PRIN 2022 PNRR grant under the National Recovery and Resilience Plan (PNRR): project P2022ZLW4T ”Next-generation computing and data technologies to probe the cosmic metal content”. LX acknowledges support from the National Natural Science Foundation of China (grant number 12041302), the Ministry of Science and Technology of China (grant No. 2020SKA0110100).
\end{acknowledgements}

%%%%%%%%%%%%%%%%%%%%%%%%%%%%%%%%%%%%%%%%%%%%%%%%%%%%%%%%%%%%%%%%%%%%%%%%%%%%%%%%%%%%%%%%%%%%%%%%%%%%%%%%%%%%%%%%%%%%%%%%%%%%%%%%%%%%%%%%%%%%%%%%%%%%%%%%%%%%%%%%%%%%%%%%%
% Bibliography
\bibliographystyle{aa}
\bibliography{bibliography}

@ARTICLE{Cantarella25,
   author = {{Cantarella}, S. and {De Lucia}, G. and {Fontanot}, F. and {Hirschmann}, H. and {Xie}, L. and {et al.,}},
  journal = {A\&A, https://arxiv.org/pdf/2511.03787},
     year = {2025},
    pages = {}}

@ARTICLE{Parente22,
   author = {{Parente}, M. and {Ragone-Figueroa}, C. and {Granato}, G. L. and {Borgani}, S. and {et al.,}},
  journal = {MNRAS},
     year = {2022},
   volume = 515,
    pages = {2053}}

@ARTICLE{Madau14,
   author = {{Madau}, P. and {Dickinson}, M.},
  journal = {ARA\&A},
     year = {2014},
   volume = 52,
    pages = {415}}

@ARTICLE{Berta25,
   author = {{Berta}, S. and {Lagache}, G. and {Beelen}, A. and {Adam}, R. and {et al.,}},
  journal = {A\&A},
     year = {2025},
   volume = 696,
    pages = {33}}

@ARTICLE{Beeston24,
   author = {{Beeston}, R. A. and {Gomez}, H. L. and {Dunne}, L. and {Maddox},  S. and {et al.,}},
  journal = {MNRAS},
     year = {2024},
   volume = 535,
    pages = {3162}}

@ARTICLE{Traina24,
   author = {{Traina}, A. and {Magnelli}, B. and {Gruppioni}, C. and {Delvecchio},  I. and {et al.,}},
  journal = {A\&A},
     year = {2024},
   volume = 690,
    pages = {10}}

@ARTICLE{Vlahakis05,
   author = {{Vlahakis}, C. and {Dunne}, L. and {Eales}, S.},
  journal = {MNRAS},
     year = {2005},
   volume = 364,
    pages = {1253}}

@ARTICLE{Hirschmann23,
   author = {{Hirschmann}, M. and {Charlot}, S. and {Somerville}, R. S.},
  journal = {MNRAS},
     year = {2023},
   volume = 526,
    pages = {3504}}

@ARTICLE{Kewley08,
   author = {{Kewley}, L. J. and {Ellison}, S. L.},
  journal = {ApJ},
     year = {2008},
   volume = 681,
    pages = {1183}}

@ARTICLE{Fontanot21,
   author = {{Fontanot}, F. and {Calabrò}, A. and {Talia}, M. and {Mannucci},  F. and {et al.,}},
  journal = {MNRAS},
     year = {2021},
   volume = 504,
    pages = {4481}}

@ARTICLE{daCunha15,
   author = {{da Cunha}, E. and {Walter}, F. and {Smail}, I. R. and {Swinbank}, A. M. and {et al.,}},
  journal = {ApJ},
     year = {2015},
   volume = 806,
    pages = {110}}

@ARTICLE{Beeston18,
   author = {{Beeston}, R. A. and {Wright}, A. H. and {Maddox}, S. and {Gomez}, H. L. and {et al.,}},
  journal = {MNRAS},
     year = {2018},
   volume = 479,
    pages = {1077}}

@ARTICLE{Nersesian19,
   author = {{Nersesian}, A. and {Xilouris}, E. M. and {Bianchi}, S. and {Galliano}, F. and {et al.,}},
  journal = {A\&A},
     year = {2019},
   volume = 624,
    pages = {A80}}

@ARTICLE{Kannan22,
   author = {{Kannan}, R. and {Smith}, A. and {Garaldi}, E. and {Shen}, X. and {et al.,}},
  journal = {MNRAS},
     year = {2022},
   volume = 514,
    pages = {3857}}

@ARTICLE{Osman25,
   author = {{Osman}, O. and {De Lucia} G. and {et al.,}},
  journal = {in prep},
     year = {2025},
    pages = {}}

@ARTICLE{Faisst25,
   author = {{Faisst}, A. and {Liu} L-J. and  {et al.,}},
  journal = {arxiv.orgxxxxx},
     year = {2025},
    pages = {}}

@ARTICLE{Nanni25,
   author = {{Nanni}, A. and {Romano}, M. and {Donevski}, D. and {Witstok}, J. and {et al.,}},
  journal = {ApJ},
     year = {2025},
   volume = 988,
    pages = {8}}

@ARTICLE{Calura08,
   author = {{Calura}, F. and {Pipino}, A. and {Matteucci}, F.},
  journal = {A\&A},
     year = {2008},
   volume = 479,
    pages = {669}}

@ARTICLE{Gioannini17,
   author = {{Gioannini}, L. and {Matteucci}, F. and {Vladilo}, G. and {Calura}, F.},
  journal = {MNRAS},
     year = {2017},
   volume = 464,
    pages = {985}}

@ARTICLE{Dayal22,
   author = {{Dayal}, P. and {Ferrara}, A. and {Sommovigo}, L. and {Bouwens}, R. and {et al.,}},
  journal = {MNRAS},
     year = {2022},
   volume = 512,
    pages = {989}}

@ARTICLE{Algera25,
   author = {{Algera}, H. S. B. and {Rowland}, L. and {Stefanon}, M. and {Palla}, M. and {et al.,}},
  journal = {https://arxiv.org/pdf/2501.10508},
     year = {2025},
    pages = {}}

@ARTICLE{Murray10,
   author = {{Murray}, N. and {Rahman}, M.},
  journal = {ApJ},
     year = {2010},
   volume = 709,
    pages = {424}}

@ARTICLE{Gutkin16,
   author = {{Gutkin}, J. and {Charlot}, S. and {Bruzual}, G.},
  journal = {MNRAS},
     year = {2016},
   volume = 462,
    pages = {1757}}

@ARTICLE{Groves04,
   author = {{Groves}, B. A. and {Dopita}, M. A. and {Sutherland}, R. S.},
  journal = {ApJS},
     year = {2004},
   volume = 153,
    pages = {75}}

@ARTICLE{Shivaei24,
   author = {{Shivaei}, I. and {Alberts}, S. and {Florian}, M. and {Rieke}, G. and {et al.,}},
  journal = {A\&A},
     year = {2024},
   volume = 690,
    pages = {17}}

@ARTICLE{Tsai95,
   author = {{Tsai}, J. C. and {Mathews}, W. G.},
  journal = {ApJ},
     year = {1995},
   volume = 448,
    pages = {84}}

@ARTICLE{Fontanot24,
   author = {{Fontanot}, F. and {La Barbera}, F. and {De Lucia}, G. and {Cecchi}, R. and {et al.,}},
  journal = {A\&A},
     year = {2024},
   volume = 686,
    pages = {13}}

@ARTICLE{Chieffi04,
   author = {{Chieffi}, A. and {Limongi}, M.},
  journal = {ApJ},
     year = {2004},
   volume = 608,
    pages = {405}}

@ARTICLE{Thielemann03,
   author = {{Thielemann}, F. -K.},
  journal = {in Hillebrandt W., Leibundgut B., eds, From
Twilight to Highlight: The Physics of Supernovae Supernova Nucle-
osynthesis and Galactic Evolution. Springer-Verlag, Berlin, pp. 331},
     year = {2003},
    pages = {331}}

@ARTICLE{Karakas10,
   author = {{Karakas}, A. I.},
  journal = {MNRAS},
     year = {2010},
   volume = 403,
    pages = {1413}}

@ARTICLE{Chabrier03,
   author = {{Chabrier}, G.},
  journal = {PASP},
     year = {2003},
   volume = 115,
    pages = {763}}

@ARTICLE{Fontanot20,
   author = {{Fontanot}, F. and {De Lucia}, G. and {Hirschmann}, H. and {Xie}, L. and {et al.,}},
  journal = {MNRAS},
     year = {2020},
   volume = 496,
    pages = {3943}}

@ARTICLE{Xie20,
   author = {{Xie}, L. and {De Lucia}, G. and {Hirschmann}, M. and {Fontanot}, F.},
  journal = {MNRAS},
     year = {2020},
   volume = 498,
    pages = {4327}}

@ARTICLE{DeLucia07,
   author = {{De Lucia}, G. and {Blaizot}, J.},
  journal = {MNRAS},
     year = {2007},
   volume = 375,
    pages = {2}}

@ARTICLE{Guo13,
   author = {{Guo}, Q. and {White}, S. and {Angulo}, R. E. and {Henriques}, B. and {et al.,}},
  journal = {MNRAS},
     year = {2013},
   volume = 428,
    pages = {1351}}

@ARTICLE{Wang08,
   author = {{Wang}, J. and {De Lucia}, G. and {Kitzbichler}, M. G. and {White}, S. D. M.},
  journal = {MNRAS},
     year = {2008},
   volume = 384,
    pages = {1301}}

@ARTICLE{Bennett13,
   author = {{Bennett}, C. L. and {Larson}, D. and {Weiland}, J. L. and {Jarosik}, N. and {et al.,}},
  journal = {ApJS},
     year = {2013},
   volume = 208,
    pages = {20}}

@ARTICLE{Springel05,
   author = {{Springel}, V. and {White}, S. D. M. and {Jenkins}, A. and {Frenk}, C. S. and {et al.,}},
  journal = {Nature},
     year = {2005},
   volume = 435,
    pages = {629}}

@ARTICLE{Boylan09,
   author = {{Boylan-Kolchin}, M. and {Springel}, V. and {White}, S. D. M. and {Jenkins}, A. and {et al.,}},
  journal = {MNRAS},
     year = {2009},
   volume = 398,
    pages = {1150}}

@ARTICLE{DeLucia14,
   author = {{De Lucia}, G. and {Tornatore}, L. and {Frenk}, C. S. and {Helmi}, A. and {et al.,}},
  journal = {MNRAS},
     year = {2014},
   volume = 445,
    pages = {970}}

@ARTICLE{Fontanot25,
   author = {{Fontanot}, F. and {De Lucia}, G. and {Xie}, L. and {Hirschmann}, H. and {et al.,}},
  journal = {A\&A},
     year = {2025},
   volume = 699,
    pages = {13}}

@ARTICLE{DeLucia24,
   author = {{De Lucia}, G. and {Fontanot}, F. and {Xie}, L. and {Hirschmann}, H.},
  journal = {A\&A},
     year = {2024},
   volume = 687,
    pages = {17}}

@ARTICLE{Xie17,
   author = {{Xie}, L. and {De Lucia}, G. and {Hirschmann}, M. and {Fontanot}, F. and {et al.,}},
  journal = {MNRAS},
     year = {2017},
   volume = 469,
    pages = {968}}

@ARTICLE{Hirschmann16,
   author = {{Hirschmann}, M. and {De Lucia}, G. and {Fontanot}, F.},
  journal = {MNRAS},
     year = {2016},
   volume = 461,
    pages = {1760}}

@ARTICLE{Watson15,
   author = {{Watson}, D. and {Christensen}, L. and {Knudsen}, K. K. and {Richard}, J. and {et al.,}},
  journal = {Nature},
     year = {2015},
   volume = 519,
    pages = {327}}

@ARTICLE{Riechers14,
   author = {{Riechers}, D. A. and {Carilli}, C. L. and {Capak}, P. L. and {Scoville}, N. Z. and {et al.,}},
  journal = {ApJ},
     year = {2014},
   volume = 796,
    pages = {84}}

@ARTICLE{Parente23,
   author = {{Parente}, P. and {Ragone-Figueroa}, C. and {Granato}, G. L. and {Lapi}, A. and {et al.,}},
  journal = {MNRAS},
     year = {2023},
   volume = 521,
    pages = {6105}}

@ARTICLE{Yates24,
   author = {{Yates}, R. M. and {Hendriks}, D. and {Vijayan}, A. P. and {Izzard}, R. G. and {et al.,}},
  journal = {MNRAS},
     year = {2024},
   volume = 527,
    pages = {6292}}

@ARTICLE{Clark15,
   author = {{Clark}, C. J. R. and {Dunne}, L. and {Gomez}, H. L. and {Maddox}, S. and {et al.,}},
  journal = {MNRAS},
     year = {2015},
   volume = 452,
    pages = {397}}

@ARTICLE{Rowlands14,
   author = {{Rowlands}, K. and {Dunne}, L. and {Dye}, S. and {Aragon-Salamanca}, A. and {et al.,}},
  journal = {MNRAS},
     year = {2014},
   volume = 441,
    pages = {1017}}

@ARTICLE{Popping22,
   author = {{Popping}, G. and {Péroux}, C.},
  journal = {MNRAS},
     year = {2022},
   volume = 513,
    pages = {1531}}

@ARTICLE{Wiseman17,
   author = {{Wiseman}, P. and {Schady}, P. and {Bolmer}, J. and {Krühler}, T. and {et al.,}},
  journal = {A\&A},
     year = {2017},
   volume = 599,
    pages = {23}}

@ARTICLE{Sparre14,
   author = {{Sparre}, M. and {Hartoog}, O. E. and {Krühler}, T. and {Fynbo}, J. P. U. and {et al.,}},
  journal = {ApJ},
     year = {2014},
   volume = 785,
    pages = {150}}

@ARTICLE{Peroux20,
   author = {{Péroux}, C. and {Howk}, J. C.},
  journal = {ARA\&A},
     year = {2020},
   volume = 58,
    pages = {363}}

@ARTICLE{Zafar13,
   author = {{Zafar}, T. and {Watson}, D.},
  journal = {A\&A},
     year = {2013},
   volume = 560,
    pages = {A26}}

@ARTICLE{DeCia16,
   author = {{De Cia}, A. and {Ledoux}, C. and {Mattsson}, L. and {Petitjean}, P. and {et al.,}},
  journal = {A\&A},
     year = {2016},
   volume = 596,
    pages = {A97}}

@ARTICLE{DeCia13,
   author = {{De Cia}, A. and {Ledoux}, C. and {Savaglio}, S. and {Schady}, P. and {et al.,}},
  journal = {A\&A},
     year = {2013},
   volume = 560,
    pages = {A88}}

@ARTICLE{Magrini11,
   author = {{Magrini}, L. and {Bianchi}, S. and {Corbelli}, E. and {Cortese}, L. and {et al.,}},
  journal = {A\&A},
     year = {2011},
   volume = 535,
    pages = {A13}}

@ARTICLE{Galametz11,
   author = {{Galametz}, M. and {Madden}, S. C. and {Galliano}, F. and {Hony}, S. and {et al.,}},
  journal = {A\&A},
     year = {2011},
   volume = 532,
    pages = {A56}}

@ARTICLE{Draine07b,
   author = {{Draine}, B. T. and {Dale}, D. A. and {Bendo}, G. and {Gordon}, K. D. and {et al.,}},
  journal = {ApJ},
     year = {2007},
   volume = 663,
    pages = {866}}

@ARTICLE{James02,
   author = {{James}, A. and {Dunne}, L. and {Eales}, S. and {Edmunds}, M. G.},
  journal = {MNRAS},
     year = {2002},
   volume = 335,
    pages = {753}}

@ARTICLE{Pozzi20,
   author = {{Pozzi}, F. and {Calura}, F. and {Zamorani}, G. and {Delvecchio}, I. and {et al.,}},
  journal = {MNRAS},
     year = {2020},
   volume = 491,
    pages = {5073}}

@ARTICLE{Clemens13,
   author = {{Clemens}, M. S. and {Negrello}, M. and {De Zotti}, G. and {Gonzalez-Nuevo}, J. and {et al.,}},
  journal = {MNRAS},
     year = {2013},
   volume = 433,
    pages = {695}}

@ARTICLE{Dunne11,
   author = {{Dunne}, L. and {Gomez}, H. L. and {da Cunha}, E. and {Charlot}, S. and {et al.,}},
  journal = {MNRAS},
     year = {2011},
   volume = 417,
    pages = {1510}}

@ARTICLE{Eales09,
   author = {{Eales}, S. and {Chapin}, E. L. and {Devlin}, M. J. and {Dye}, S. and {et al.,}},
  journal = {ApJ},
     year = {2009},
   volume = 707,
    pages = {1779}}

@ARTICLE{Dunne03,
   author = {{Dunne}, L. and {Eales}, S. A. and {Edmunds}, M. G.},
  journal = {MNRAS},
     year = {2003},
   volume = 341,
    pages = {589}}

@ARTICLE{DeVis19,
   author = {{De Vis}, P. and {Jones}, A. and {Viaene}, S. and {Casasola}, V. and {et al.,}},
  journal = {A\&A},
     year = {2019},
   volume = 623,
    pages = {A5}}

@ARTICLE{Casey12,
   author = {{Casey}, C. M.},
  journal = {MNRAS},
     year = {2012},
   volume = 425,
    pages = {3094}}

@ARTICLE{daCunha10,
   author = {{da Cunha}, E. and {Eminian}, C. and {Charlot}, S. and {Blaizot}, J.},
  journal = {MNRAS},
     year = {2010},
   volume = 403,
    pages = {1894}}

@ARTICLE{Santini14,
   author = {{Santini}, P. and {Maiolino}, R. and {Magnelli}, B. and {Lutz}, D. and {et al.,}},
  journal = {A\&A},
     year = {2014},
   volume = 562,
    pages = {A30}}

@ARTICLE{Corbelli12,
   author = {{Corbelli}, E. and {Bianchi}, S. and {Cortese}, L. and {Giovanardi}, C. and {et al.,}},
  journal = {A\&A},
     year = {2012},
   volume = 542,
    pages = {A32}}

@ARTICLE{Bigiel08,
   author = {{Bigiel}, F. and {Leroy}, A. and {Walter}, F. and {Brinks}, E. and {et al.,}},
  journal = {AJ},
     year = {2008},
   volume = 136,
    pages = {2846}}

@ARTICLE{osman20b,
   author = {{Osman}, O. and {Bekki}, K. and {Cortese}, L.},
  journal = {MNRAS},
     year = {2020},
   volume = 498,
    pages = {2075}}

@ARTICLE{Andersen11,
   author = {{Andersen}, M. and {Rho}, J. and {Reach}, W. T. and {Hewitt}, J. W. and {Bernard}, J. P.},
  journal = {ApJ},
     year = 2011,
   volume = 742,
    pages = {7}}

@ARTICLE{Aoyama17,
   author = {{Aoyama}, S. and   {Hou}, K. and {Shimizu}, I. and  {Hirashita}, H. and {Todoroki}, K. and {Choi}, J. and {Nagamine}, K.},
  journal = {MNRAS},
     year = 2017,
   volume = 466,
    pages = {105}}

@ARTICLE{Aoyama18,
   author = {{Aoyama},  S. and   {Hou},  K. and   {Hirashita},  H. and {Nagamine},  K. and  {Shimizu},  I.},
  journal = {MNRAS},
     year = 2018,
   volume = 478,
    pages = {4905}}

@ARTICLE{Asano13a,
   author = {{Asano}, R. S. and  {Takeuchi}, T. T. {Hirashita}, H. and {Inoue}, A. K.},
  journal = {Earth, Planets and Space},
     year = {2013},
   volume = 65,
    pages = {213}}

@ARTICLE{Barlow78,
   author = {{Barlow}, M. J.},
  journal = {MNRAS},
     year = 1978,
   volume = 183,
    pages = {367}}

@ARTICLE{Bekki13,
   author = {{Bekki}, K.},
  journal = {MNRAS},
     year = 2013,
   volume = 432,
    pages = {2298}}

@ARTICLE{Bekki15,
   author = {{Bekki}, K.},
  journal = {MNRAS},
     year = 2015,
   volume = 449,
    pages = {1625}}

@ARTICLE{Bertoldi03,
   author = {{Bertoldi}, F. and {Carilli}, C. L. and {Cox}, P. and {Fan}, X. and {Strauss}, M. A. and {Beelen}, A. and {Omont}, A. and {Zylka}, R.},
  journal = {A\&A},
     year = 2003,
   volume = 406,
    pages = {L55}}

@ARTICLE{Bianchi07,
   author = {{Bianchi}, S. and {Schneider}, R.},
  journal = {MNRAS},
     year = 2007,
   volume = 378,
    pages = {973}}

@ARTICLE{Cazaux04,
   author = {{Cazaux}, S. and {Tielens}, A. G. G. M.},
  journal = {ApJ},
     year = 2004,
   volume = 604,
    pages = {222}}

@ARTICLE{Draine03,
   author = {{Draine}, B. T.},
  journal = {ARA\&A},
     year = 2003,
   volume = 41,
    pages = {241}}

@ARTICLE{Draine07,
   author = {{Draine}, B. T.},
  journal = {ApJ},
     year = 2007,
   volume = 663,
    pages = {894}}

@ARTICLE{Draine11,
   author = {{Draine}, B. T.},
  journal = {Physics of the Interstellar and Intergalactic Medium by Bruce T. Draine. Princeton University Press, 2011. ISBN: 978-0-691-12214-4},
     year = 2011}

@ARTICLE{Dulieu13,
   author = {{Dulieu},  F. and  {Congiu},  E. and  {Noble},  J. and  {Baouche},  S. and  {Chaabouni},  H. and  {Moudens},  A. and {Minissale}, M. and {Cazaux}, S.},
  journal = {Sci. Rep.},
     year = 2013,
   volume = 3,
    pages = {1338}}

@ARTICLE{Dwek80,
   author = {{Dwek}, E. and {Scalo}, J. M.},
  journal = {ApJ},
     year = 1980,
   volume = 239,
    pages = {193}}

@ARTICLE{Dwek98,
   author = {{Dwek}, E.},
  journal = {ApJ},
     year = 1998,
   volume = 501,
    pages = {643}}

@ARTICLE{Ferrarotti06,
   author = {{Ferrarotti}, A. D. and {Gail}, H. P.},
  journal = {A\&A},
     year = 2006,
   volume = 553,
    pages = {576}}

@ARTICLE{Forbes16,
   author = {{Forbes}, J. C. and {Krumholz}, M. R. and {Goldbaum}, N. J. and {Dekel}, A.},
  journal = {Nature},
     year = 2016,
   volume = 535,
    pages = {523}}

@ARTICLE{Fukui10,
   author = {{Fukui}, Y. and {Kawamura}, A.},
  journal = {ARA\&A},
     year = 2010,
   volume = 48,
    pages = {547}}

@ARTICLE{Galliano18,
   author = {{Galliano}, F. and {Galametz}, M. and {Jones}, A. P.},
  journal = {ARA\&A},
     year = 2018,
   volume = 56,
    pages = {673}}

@ARTICLE{Ginolfi18,
   author = {{Ginolfi}, M. and {Graziani}, L. and {Schneider}, R. and {Marassi}, S. and {Valiante}, R. and {Dell’Agli}, F. and {Ventura}, P. and {Hunt}, L. K.},
  journal = {MNRAS},
     year = 2018,
   volume = 473,
    pages = {4538}}

@ARTICLE{Gjergo18,
   author = {{Gjergo}, E. and {Granato}, G. L. and {Murante}, G. and {Ragone-Figueroa}, C. and {Tornatore}, L. and {Borgani}, S.},
  journal = {MNRAS},
     year = 2018,
   volume = 479,
    pages = {2588}}

@ARTICLE{Hirashita11,
   author = {{Hirashita}, H. and {Kuo}, T.},
  journal = {MNRAS},
     year = 2011,
   volume = 416,
    pages = {1340}}

@ARTICLE{Hirashita19,
   author = {{Hirashita}, H. and {Aoyama}, S.},
  journal = {MNRAS},
     year = 2019,
   volume = 482,
    pages = {2555}}

@ARTICLE{Hou17,
   author = {{Hou}, K. and {Hirashita}, H. and {Nagamine}, K. and {Aoyama}, S. and {Shimizu}, I.},
  journal = {MNRAS},
     year = 2017,
   volume = 469,
    pages = {870}}

@ARTICLE{Hu17,
   author = {{Hu}, C.-Y. and {Naab}, T. and {Glover}, S. C. O. and {Walch}, S. and {Clark}, P. C.},
  journal = {MNRAS},
     year = 2017,
   volume = 471,
    pages = {2151}}

@ARTICLE{Jones94,
   author = {{Jones}, A. P. and {Tielens}, A. G. G. M. and {Hollenbach}, D. J. and {McKee}, C. F.},
  journal = {ApJ},
     year = 1994,
   volume = 433,
    pages = {797}}

@ARTICLE{Kuo12,
   author = {{Kuo}, T.-M. and {Hirashita}, H.},
  journal = {MNRAS},
     year = 2012,
   volume = 424,
    pages = {L34}}

@ARTICLE{Mattsson11,
   author = {{Mattsson}, L.},
  journal = {MNRAS},
     year = 2011,
   volume = 414,
    pages = {781}}

@ARTICLE{Mckee89,
   author = {{McKee}, C.},
  journal = {in Allamandola L. J., Tielens A. G. G. M., eds, Proc. IAU Symp. 135, Interstellar Dust. Kluwer, Dordrecht, 431},
     year = 1989}

@ARTICLE{McKinnon16,
   author = {{McKinnon}, R. and {Torrey}, P. and {Vogelsberger}, M.},
  journal = {MNRAS},
     year = 2016,
   volume = 457,
    pages = {3775}}

@ARTICLE{McKinnon18,
   author = {{McKinnon}, R. and {Torrey}, P. and {Vogelsberger}, M.},
  journal = {MNRAS},
     year = 2018,
   volume = 478,
    pages = {2851}}

@ARTICLE{Micelotta18,
   author = {{Micelotta}, E. R. and {Matsuura}, M. and {Sarangi}, A.},
  journal = {Space Sci. Rev.},
     year = 2018,
   volume = 214,
    pages = {58}}

@ARTICLE{Nozawa06,
   author = {{Nozawa}, T. and {Kozasa}, T. and  {Habe}, A.},
  journal = {ApJ},
     year = 2006,
   volume = 648,
    pages = {435}}

@ARTICLE{Relano18,
   author = {{Rela\~{n}o}, M. and {De Looze}, I. and {Kennicutt}, R. C. and {Lisenfeld}, U.  and {et al.,}},
  journal = {A\&A},
     year = 2018,
   volume = 613,
    pages = {A43}}

@ARTICLE{Remy14,
   author = {{R\'{e}my-Ruyer}, A. and {Madden}, S. C. and {Galliano}, F. and {Galametz}, M. and {et al.,}},
  journal = {A\&A},
     year = 2014,
   volume = 563,
    pages = {A31}}

@ARTICLE{Sargent10,
   author = {{Sargent}, B. A. and {Srinivasan}, S. and {Meixner}, M. and {Kemper}, F. and {et al.,}},
  journal = {ApJ},
     year = 2010,
   volume = 716,
    pages = {878}}

@ARTICLE{Savage96,
   author = {{Savage}, B. D. and {Sembach},  K. R.},
  journal = {ARA\&A},
     year = 1996,
   volume = 34,
    pages = {279}}

@ARTICLE{Smith12,
   author = {{Smith}, D. J. B. and {Dunne}, L. and {da Cunha}, E. and {Rowlands}, K. and {et al.,}},
  journal = {MNRAS},
     year = 2012,
   volume = 427,
    pages = {703}}

@ARTICLE{Srinivasan10,
   author = {{Srinivasan}, S. and {Sargent}, B. A. and {Matsuura}, M. and {Meixner}, M. and {et al.,}},
  journal = {A\&A},
     year = 2010,
   volume = 524,
    pages = {A49}}

@ARTICLE{Ventura12a,
   author = {{Ventura}, P. and {di Criscienzo}, M. and {Schneider}, R. and {Carini}, R. and {et al.,}},
  journal = {MNRAS},
     year = {2012},
   volume = 420,
    pages = {1442}}

@ARTICLE{Wakelam17,
   author = {{Wakelam}, V. and {Bron}, E. and {Cazaux}, S. and {Dulieu}, F. and {et al.,}},
  journal = {Molecular Astrophys},
     year = 2017,
   volume = 9,
    pages = {1}}

@ARTICLE{Yamasawa11,
   author = {{Yamasawa}, D. and  {Habe}, A. and  {Kozasa}, T. and   {Nozawa}, T. and   {Hirashita}, H.  and {Umeda}, H. and {Nomoto}, K.},
  journal = {ApJ},
     year = 2011,
   volume = 734,
    pages = {44}}

@ARTICLE{Zhukovska16,
   author = {{Zhukovska}, S. and  {Dobbs}, C. and  {Jenkins}, E. B. and   {Klessen}, R. S.},
  journal = {ApJ},
     year = 2016,
   volume = 831,
    pages = {147}}

@ARTICLE{Zhukovska18,
   author = {{Zhukovska}, S. and {Henning}, T. and {Dobbs}, C.},
  journal = {ApJ},
     year = 2018,
   volume = 857,
    pages = {94}}

@ARTICLE{Issa90,
   author = {{Issa}, M. R. and {MacLaren}, I. and  {Wolfendale}, A. W.},
  journal = {A\&A},
     year = 1990,
   volume = 236,
    pages = {237}}

@ARTICLE{Lisenfeld98,
   author = {{Lisenfeld}, U. and  {Ferrara}, A.},
  journal = {ApJ},
     year = 1998,
   volume = 496,
    pages = {145}}

@ARTICLE{Hirashita02,
   author = {{Hirashita}, H. and {Tajiri}, Y. Y. and {Kamaya}, H.},
  journal = {A\&A},
     year = 2002,
   volume = 388,
    pages = {439}}

@ARTICLE{Cortese12,
   author = {{Cortese}, L. and {Ciesla}, L. and {Boselli}, A. and {Bianchi}, S. and {et al.,}},
  journal = {A\&A},
     year = 2012,
   volume = 540,
    pages = {A52}}

@ARTICLE{Grossi15,
   author = {{Grossi}, M. and {Hunt}, L. K. and {Madden}, S. C. and {Hughes}, T. M. and {et al.,}},
  journal = {A\&A},
     year = 2015,
   volume = 574,
    pages = {A126}}

@ARTICLE{Hirashita13,
   author = {{Hirashita}, H.},
  journal = {Proceedings of The Life Cycle of Dust in the Universe: Observations, Theory, and Laboratory Experiments (LCDU2013), p. 27. Available at: http://pos.sissa.it/cgi-bin/reader/conf.cgi?confid=207},
     year = 2013,}

@ARTICLE{Hirashita99,
   author = {{Hirashita}, H.},
  journal = {ApJ},
     year = 1999,
   volume = 510,
    pages = {L99}}

@ARTICLE{Asano13,
   author = {{Asano}, R. S. and  {Takeuchi}, T. T. and  {Hirashita}, H.  and  {Nozawa}, T.},
  journal = {MNRAS},
     year = 2013,
   volume = 432,
    pages = {637}}

@ARTICLE{Bakes94,
   author = {{Bakes}, E. L. O. and {Tielens}, A. G. G. M.},
  journal = {ApJ},
     year = 1994,
   volume = 427,
    pages = {822}}

@ARTICLE{Draine78,
   author = {{Draine}, B. T.},
  journal = {ApJSS},
     year = 1978,
   volume = 36,
    pages = {595}}

@ARTICLE{Hill18,
   author = {{Hill}, A. S. and {Mac Low}, M.-M. and {Gatto}, A. and {Ibanez-Mejia}, J. C.},
  journal = {ApJ},
     year = 2018,
   volume = 861,
    pages = {55}}

@ARTICLE{Ingalls02,
   author = {{Ingalls}, J. G. and {Reach}, W. T. and {Bania}, T. M.},
  journal = {ApJ},
     year = 2002,
   volume = 579,
    pages = {289}}

@ARTICLE{Watson72,
   author = {{Watson}, W. D.},
  journal = {ApJ},
     year = 1972,
   volume = 176,
    pages = {103}}

@ARTICLE{Weingartner01,
   author = {{Weingartner}, J. C. and {Draine}, B. T.},
  journal = {ApJSS},
     year = 2001,
   volume = 134,
    pages = {263}}

@ARTICLE{Wolfire95,
   author = {{Wolfire}, M. G. and {Hollenbach},D. and {McKee}, C. F. and {Tielens}, A. G. G. M. and {Bakes}, E.L. O.},
  journal = {ApJ},
     year = 1995,
   volume = 443,
    pages = {152}}

@ARTICLE{Wolfire03,
   author = {{Wolfire}, M. G. and {Mckee}, C. F. and {Tielens}, A. G. G. M.},
  journal = {ApJ},
     year = 2003,
   volume = 587,
    pages = {278}}

@ARTICLE{Bianchi05,
   author = {{Bianchi}, S. and {Ferrara}, A.},
  journal = {MNRAS},
     year = 2005,
   volume = 358,
    pages = {379}}

@ARTICLE{Ferrara91,
   author = {{Ferrara}, A. and {Ferrini}, F. and {Barsella}, B. and {Franco}, J.},
  journal = {ApJ},
     year = 1991,
   volume = 381,
    pages = {137}}

@ARTICLE{Li19,
   author = {{Li}, Q. and {Narayanan}, D. and {Dave}, R.},
  journal = {MNRAS},
     year = 2019,
   volume = 490,
    pages = {1425}}

@ARTICLE{McKinnon17,
   author = {{McKinnon}, R. and {Torrey}, P. and {Vogelsberger}, M. and {Hayward}, C. C. and {Marinacci}, F.},
  journal = {MNRAS},
     year = 2017,
   volume = 468,
    pages = {1505}}

@ARTICLE{Hirashita15,
   author = {{Hirashita}, H.},
  journal = {MNRAS},
     year = 2015,
   volume = 447,
    pages = {2937}}

@ARTICLE{Aoyama20,
   author = {{Aoyama}, S. and {Hirashita}, H. and {Nagamine}, K.},
  journal = {MNRAS},
     year = 2020,
   volume = 491,
    pages = {3844}}

@ARTICLE{Granato21,
   author = {{Granato}, G. L. and {Ragone-Figueroa}, C. and {Taverna}, A. and {Silva}, L. and {et al.,}},
  journal = {MNRAS},
     year = 2021,
   volume = 503,
    pages = {511}}

@ARTICLE{Schruba11,
   author = {{Schruba}, A. and {Leroy}, A. K. and {Walter}, F. and {Bigiel}, F. and {et al.,}},
  journal = {AJ},
     year = 2011,
   volume = 142,
    pages = {37}}

@ARTICLE{Zhukovska08,
   author = {{Zhukovska}, S. and {Gail}, H.-P. and {Trieloff}, M.},
  journal = {A\&A},
     year = 2008,
   volume = 479,
    pages = {453}}

@ARTICLE{Klessen16,
   author = {{Klessen}, R. S. and {Glover}, S. C. O},
  journal = {Star Formation in Galaxy Evolution: Connecting Numerical Models to Reality, Saas-Fee Advanced Course, Volume 43. ISBN 978-3-662-47889-9. Springer-Verlag Berlin Heidelberg, p. 85},
     year = 2016,}

@ARTICLE{Planck16,
   author = {{Planck Collaboration}, {et al.,}},
  journal = {A\&A},
     year = 2016,
   volume = 594,
    pages = {A13}}

@ARTICLE{Mathis77,
   author = {{Mathis}, J. S. and {Rumpl}, W. and {Nordsieck}, K. H.},
  journal = {ApJ},
     year = 1977,
   volume = 217,
    pages = {425}}

@ARTICLE{Williams87,
   author = {Williams, D. A.},
  journal = {IN: Astrochemistry; Proceedings of the IAU Symposium, Goa, India, Dec. 3-7, 1985 (A87-47376 21-90). Dordrecht, D. Reidel Publishing Co., 1987, p. 531-536; Discussion, p. 537, 538},
     year = 1987}

@ARTICLE{Jones85,
   author = {{Jones}, A. P. and  {Williams}, D. A.},
  journal = {MNRAS},
     year = 1985,
   volume = 217,
    pages = {413}}

@ARTICLE{Demyk11,
   author = {{Demyk}, K.},
  journal = {Chemistry in Astrophysical Media, AstrOHP 2010, Observatoire de Haute-Provence, France, Edited by T. Pino; E. Dartois; EPJ Web of Conferences, Volume 18, id.03001},
     year = 2011}

@ARTICLE{Sarangi18,
   author = {{Sarangi}, A. and {Matsuura}, M. and {Micelotta}, E. R.},
  journal = {Space Sci. Rev.},
     year = 2018,
   volume = 214,
    pages = {48}}

@ARTICLE{Black87,
   author = {{Black}, J. H.},
  journal = {Interstellar Processes, Proceedings of a symposium, held at Grand Teton National Park, Wyo., July, 1986, Dordrecht: Reidel, 1987, edited by David J. Hollenbach, and Harley A. Thronson. Astrophysics and Space Science Library.},
     year = 1987,
   volume = 134,
    pages = {731}}

@ARTICLE{Glover12,
   author = {{Glover}, S. C. O. and {Clark}, P. C.},
  journal = {MNRAS},
     year = 2012,
   volume = 421,
    pages = {9}}

@ARTICLE{Nozawa03,
   author = {{Nozawa}, T. and {Kozasa}, T. and {Umeda}, H. and {Maeda}, K. and {Nomoto}, K.},
  journal = {ApJ},
     year = 2003,
   volume = 598,
    pages = {785}}

@ARTICLE{Whittet89,
   author = {{Whittet}, D.},
  journal = {Interstellar Dust: Proceedings of the 135th Symposium of the International Astronomical Union, held in Santa Clara, California, 26-30 July 1988. Edited by Louis J. Allamandola and A. G. G. M. Tielens. International Astronomical Union. Symposium no. 135, Kluwer Academic Publishers, Dordrecht, p.455},
     year = 1989}

@ARTICLE{Bianchi18,
   author = {{Bianchi}, S. and {De Vis}, P. and {Viaene}, S. and {Nersesian}, A. and {et al.,}},
  journal = {A\&A},
     year = 2018,
   volume = 620,
    pages = {21}}

@ARTICLE{Jones04,
   author = {{Jones}, A. P},
  journal = {
    Astrophysics of Dust, ASP Conference Series, Vol. 309,Proceedings of the conference held 26-30 May, 2003 in Estes Park, Colorado. Edited by Adolf N. Witt, Geoffrey C. Clayton and Bruce T. Draine., p.347},
     year = 2004}

@ARTICLE{Bocchio16,
   author = {{Bocchio}, M. and {Marassi}, S. and {Schneider}, R. and {Bianchi}, S. and {Limongi}, M. and {Chieffi}, A},
  journal = {A\&A},
     year = 2016,
   volume = 587,
    pages = {14}}

@ARTICLE{Duley78,
   author = {{Duley}, W. W. and {Millar}, T. J.},
  journal = {ApJ},
     year = 1978,
   volume = 220,
    pages = {124}}

@ARTICLE{Hirashita12,
   author = {{Hirashita}, H.},
  journal = {MNRAS},
     year = 2012,
   volume = 422,
    pages = {1263}}

@ARTICLE{Valiante09,
   author = {{Valiante}, R. and {Schneider}, R. and {Bianchi}, S. and {Andersen}, A. C.},
  journal = {MNRAS},
     year = 2009,
   volume = 397,
    pages = {1661}}

@ARTICLE{Popping17,
   author = {{Popping}, G. and {Somerville}, R. S. and {Galametz}, M},
  journal = {MNRAS},
     year = 2017,
   volume = 471,
    pages = {3152}}

@ARTICLE{Vijayan19,
   author = {{Vijayan}, A. P. and {Clay}, S. J. and {Thomas}, P. A. and {Yates}, R. M. and {Wilkins}, S. M. and {Henriques}, B. M.},
  journal = {MNRAS},
     year = 2019,
   volume = 489,
    pages = {4072}}

@ARTICLE{Triani20,
   author = {{Triani}, D. P. and {Sinha}, M. and {Croton}, D. J. and {Pacifici}, C. and {Dwek}, E.},
  journal = {MNRAS},
     year = 2020,
   volume = 493,
    pages = {2490}}

@ARTICLE{Hou19,
   author = {{Hou}, K-C. and {Aoyama}, S. and {Hirashita}, H. and {Nagamine}, K. and {Shimizu}, I.},
  journal = {MNRAS},
     year = 2019,
   volume = 485, 
    pages = {1727}}

@ARTICLE{Inoue03,
   author = {{Inoue}, A. K.},
  journal = {PASJ},
     year = 2003,
   volume = 55,
    pages = {901}}

@ARTICLE{Dorschner95,
   author = {{Dorschner}, J. and {Henning}, T.},
  journal = {ARA\&A},
     year = 1995,
   volume = 6,
    pages = {271}}

@ARTICLE{Remy15,
   author = {{R\'{e}my-Ruyer}, A. and {Madden}, S. C. and {Galliano}, F. and {Lebouteiller}, V. and {Baes}, M. and {et al.,}},
  journal = {A\&A},
     year = 2015,
   volume = 582,
    pages = {42}}

@ARTICLE{Aguirre01,
   author = {{Aguirre}, A. amd {Hernquist}, L. and {Katz}, N. and {Gardner}, J. and {Weinberg}, D.},
  journal = {ApJ},
     year = 2001,
   volume = 556,
    pages = {L11}}
%
%\appendix
\begin{appendix}
\onecolumn
%%%%%%%%%%%%%%%%%%%%%%%%%%%%%%%%%%%%%%%%%%%%%%%%%%%%%%%%%%%%%%%%%%%%%%%%%%%%%%%%%%%%%%%%%%%%%%%%%%%%%%%%%%%%%%%%%%%%%%%%%%%%%%%%%%%%%%%%%%%%%%%%%%%%%%%%%%%%%%%%%%%%%%%%%
\section{Dust destruction models}
\label{appB}

There are two different approaches that can be found in literature for estimating the destruction time scale. These employ the concepts of the mass of the ISM completely cleared of dust by SN shocks (\citealp{Valiante09}; \citealp{Popping17}; \citealp{Vijayan19}) and the mass of the ISM swept up by the shocks (\citealp{Mckee89}; \citealp{Dwek98}; \citealp{McKinnon16}; \citealp{Aoyama17}; \citealp{Li19}). In section \ref{des}, we present the model we used in most of our analysis, which is a flavor of the second approach. In this section, we present two extra models, employing flavors of the two approaches.

\subsection{Destruction time scales}
The destruction time scale for the first model adopting the first approach is given by the following equation:
\begin{equation}
\tau_{des} = \begin{array}{rcl} \frac{M_{HI+HII}}{M_{cleared} \gamma_{SN} \xi_{SN}} & yr \end{array},
\end{equation}
where $M_{HI+HII}$ is the mass of the atomic and ionized gas, $M_{cleared}$ is the mass of the ISM cleared of dust, $\gamma_{SN}$ is SNe rate, and $\xi_{SN}$ is the fraction of clustered SNe. We adopt values of $M_{cleared}$ = 980 for silicates and 600 for carbon dust (\citealp{Popping17}).

For the second model adopting the second approach, the destruction time scale is given by the following equation (same as equation \ref{des1}):
\begin{equation}
\tau_{des} = \begin{array}{rcl} \frac{M_{HI+HII}}{M_{swept} \gamma_{SN} \xi_{SN} \zeta_{SN}} & yr \end{array},
\end{equation}
where $M_{swept}$ is the mass of the ISM swept up,  and $\zeta_{SN}$ is SNe destruction efficiency. For this model, we adopt the functional form  by \cite{Yamasawa11} for $M_{swept}$ and the same parameter values as in our fiducial model.

\begin{equation}
M_{swept} = \begin{array}{rcl} 1535(n_0)^{-0.202}[(\frac{Z}{Z_\odot}) + 0.039]^{-0.298} & M_{\odot} \end{array},
\end{equation}
where Z is the gas metallicity. This is a fitting formula, and is accurate within less than 16\% for $n_0$ between 0.03 and 30 cm$^{-3}$, and $\frac{Z}{Z_\odot}$ between 10$^{-4}$ and 1. 

\subsection{Dust mass-stellar mass relation}
\begin{figure*}
\centering
    \includegraphics[width=1.\linewidth]{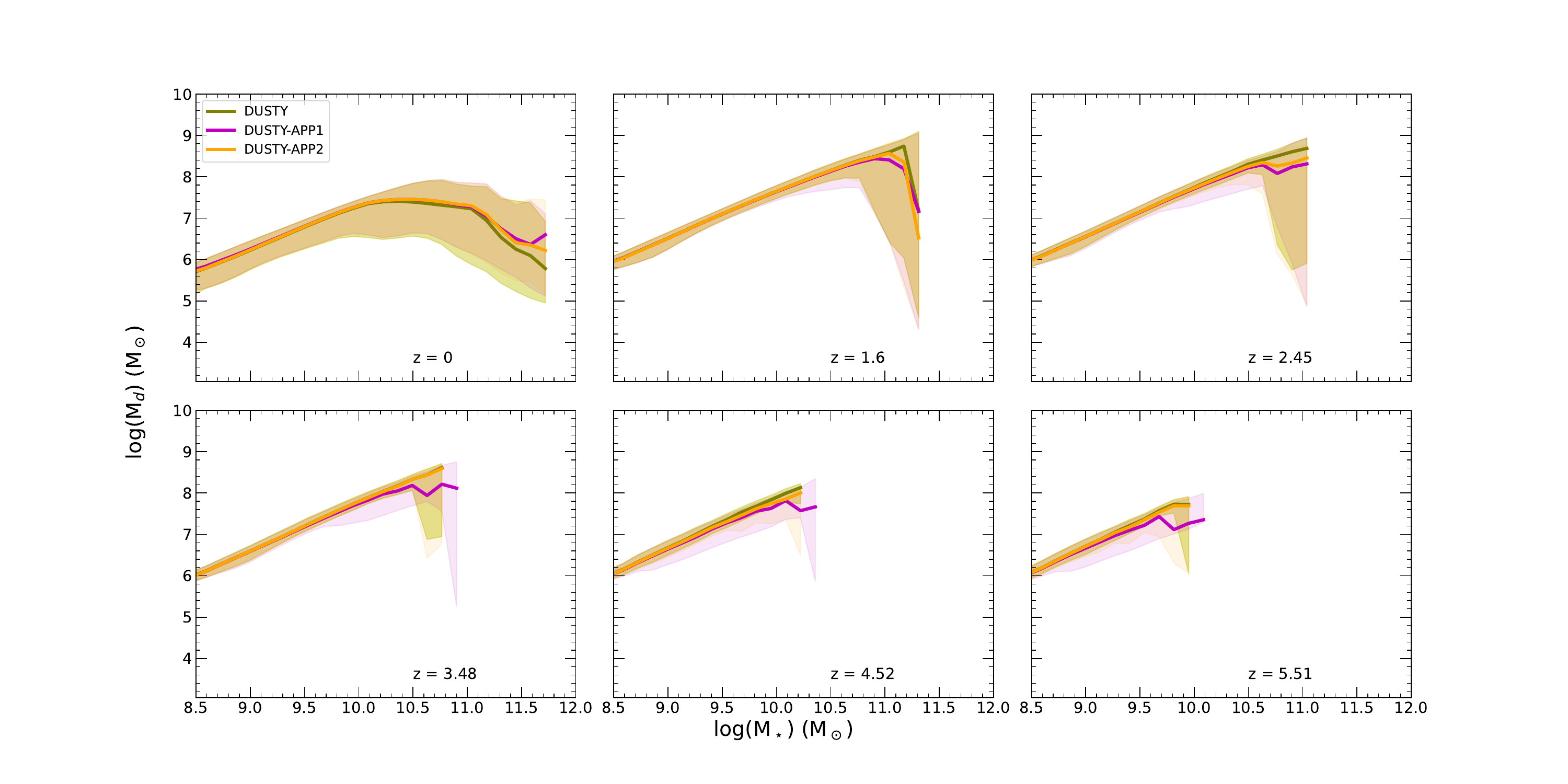}\par      
    \caption{The dust mass as a function of the stellar mass at different redshifts. Olive (DUSTY refers to the model DUSTY-GAEA), magenta and orange lines and shaded areas represent predictions of the fiducial, the model adopting the first approach, and the model adopting the second modified approach, respectively. Shaded areas represent the 16th-84th percentile region.}
\vspace{-0.3cm}
\label{fig3-}
\end{figure*}

For these test models, we only run the model using the MSI merger trees. Figure \ref{fig3-} shows predictions of the dust mass as a function of the stellar mass from z $\sim$ 0 to 5.5 in the two models presented here, together with the fiducial model. Olive, magenta and orange lines and shaded areas represent predictions of the fiducial, the model adopting the first approach, and the model adopting the second modified approach, respectively.  In the mass range resolved by the MSI merger trees, predictions of the three models are practically identical, which we attribute to the efficient growth in this mass range, making the results insensitive to the changes we implement in the destruction model. The situation could be different at low stellar masses. We will address the destruction efficiency in the low mass range together with variations in the condensation efficiency in stellar ejecta in future works.

%%%%%%%%%%%%%%%%%%%%%%%%%%%%%%%%%%%%%%%%%%%%%%%%%%%%%%%%%%%%%%%%%%%%%%%%%%%%%%%%%%%%%%%%%%%%%%%%%%%%%%%%%%%%%%%%%%%%%%%%%%%%%%%%%%%%%%%%%%%%%%%%%%%%%%%%%%%%%%%%%%%%%%%%%

\section{Dust formation and destruction rates}
\label{appA}

\begin{figure*}
\centering
    \includegraphics[width=1.\linewidth]{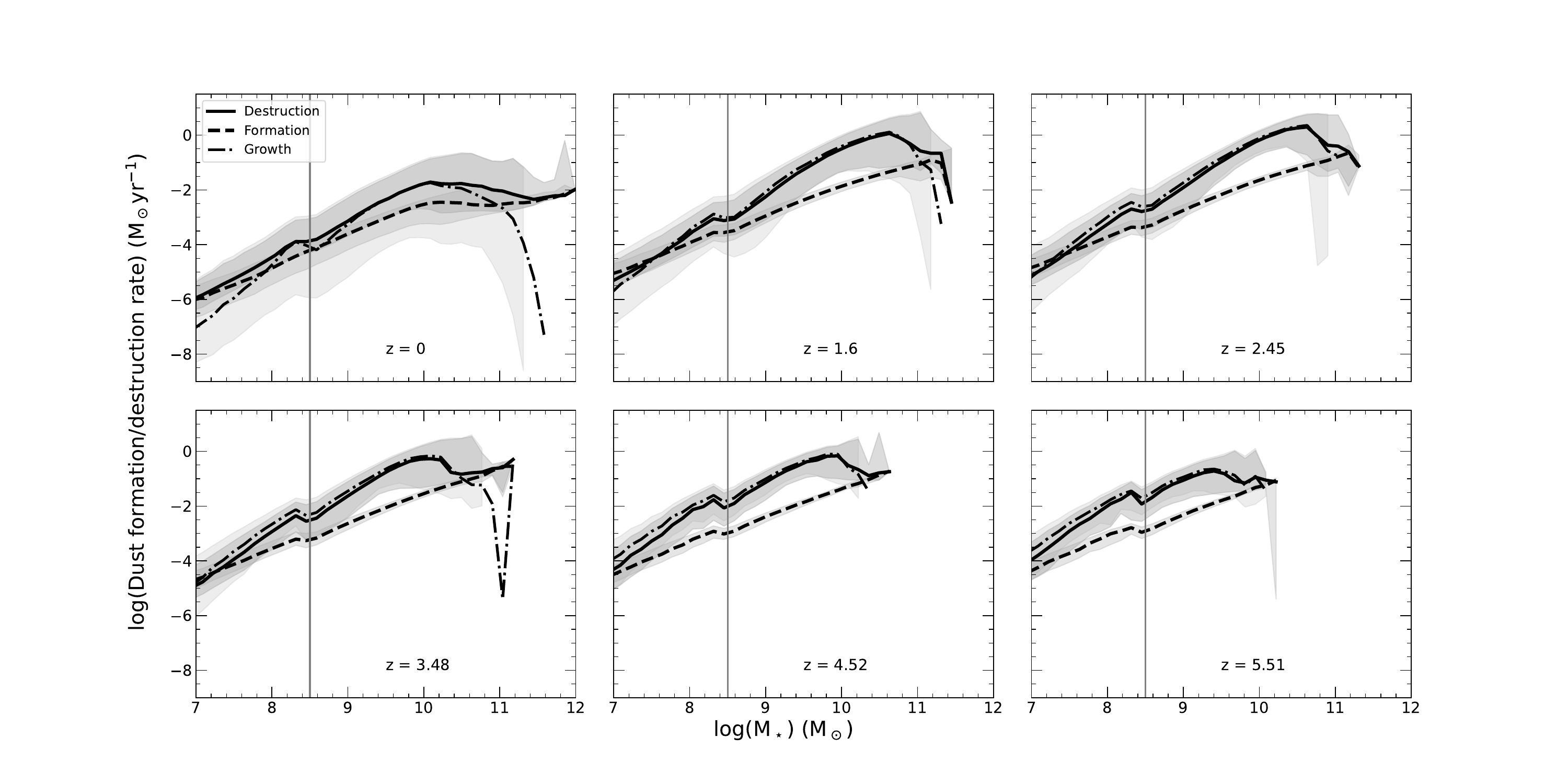}\par      
    \caption{The dust formation, destruction and growth rates as predicted by the DUSTY-GAEA-Dwek model. The solid-dashed, solid and dashed-dotted lines represent the dust formation rates by stars, destruction by SNe forward shocks, and growth in the dense ISM, respectively. Shaded areas represent the 16th-84th percentile region.}
\vspace{-0.3cm}
\label{ratesA}
\end{figure*}

Figure \ref{ratesA} presents the dust formation rates by stars (solid-dashed), destruction by SNe forward shocks (solid) and growth in the dense ISM (dashed-dotted) in the model variant DUSTY-GAEA-Dwek. As discussed in figure \ref{rates}, the rates presented here behave similarly to the rates in the model variant DUSTY-GAEA. Besides the slightly steeper correlation between the dust growth rate and stellar mass in the DUSTY-GAEA-Dwek model, this model variant also has a larger scatter compared to the DUSTY-GAEA variant.

%%%%%%%%%%%%%%%%%%%%%%%%%%%%%%%%%%%%%%%%%%%%%%%%%%%%%%%%%%%%%%%%%%%%%%%%%%%%%%%%%%%%%%%%%%%%%%%%%%%%%%%%%%%%%%%%%%%%%%%%%%%%%%%%%%%%%%%%%%%%%%%%%%%%%%%%%%%%%%%%%%%%%%%%%

\section{The dust mass function in hydrodynamical simulations}
\label{appC}

In this appendix, we compare our fiducial model predictions of the DMF to predictions from the hydrodynamical simulations by \cite{McKinnon17}, \cite{Aoyama18}, \cite{Hou19}, \cite{Li19}, and \cite{Parente22}. All these simulations explicitly model dust formation and evolution, where they include formation by stars, destruction by SNe and sputtering in the hot gas, and growth in the dense ISM. Simulations by \cite{Aoyama18}, \cite{Hou19}, and \cite{Parente22} also include shattering and coagulation of dust grains. These latter models employ the same dust growth and destruction models as those described in \cite{Hirashita11} and the formulae developed by \cite{Aoyama17}, respectively. Meanwhile,  \cite{McKinnon17} and \cite{Li19} adopted the classic prescriptions by \cite{Dwek98} and \cite{Mckee89}, respectively. Each of these simulations adopted slightly different prescriptions for dust formation by stars. For further details, we refer the reader to the original papers. Note that the biggest volume simulated by these simulations is 100 cMpc by \cite{Li19}.

In the local Universe, our model predicts a higher number density than all models at the low dust mass end ($<$ 10$^6$ M$_\odot$), while at the high mass end ($>$ 10$^{8.5}$ M$_\odot$), our model predictions are only higher than predictions by McKinnon et al. and Parente et al. In the intermediate mass range, our predictions are broadly consistent with all but the simulation by Aoyama et al. Similar trends are seen in predictions beyond the local Universe, where our model dominates the low mass end and some of the other models dominate the high mass end. The behaviour of these simulations is quite diverse, and only simulations by McKinnon et al., Li et al., and Parente et al. reproduce well the observational constraints in the local Universe. Aoyama et al. attributed their high number density at the high dust mass end to inefficient feedback at the high end of the stellar mass range, while Hou et al. argued that the inconsistency with observations is due to dust overproduction.

\begin{figure*}
\centering
    \makebox[\textwidth][c]{\includegraphics[width=1.1\linewidth]{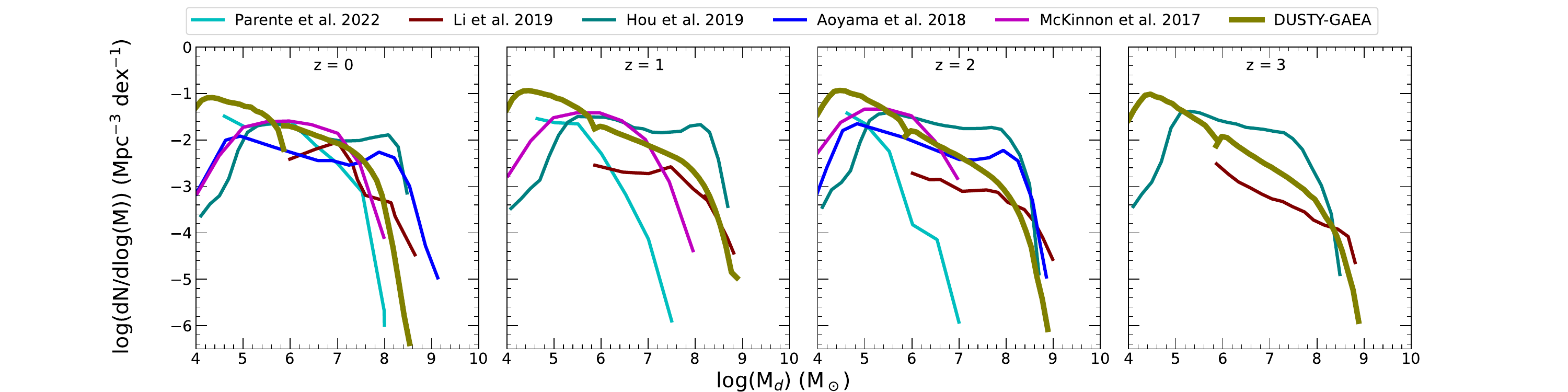}\par}     
    \caption{The dust mass function (DMF) from redshift $\sim$ 0 to 3. Solid olive lines represent predictions by DUSTY-GAEA, combining the MSI and MSII runs. Coloured solid lines represent predictions from the hydrodynamical simulations by \cite{McKinnon17} (magenta), \cite{Aoyama18} (blue), \cite{Hou19} (teal), \cite{Li19} (maroon), and \cite{Parente22} (cyan).}
\vspace{-0.3cm}
\label{DMF-hy}
\end{figure*}

\end{appendix}

\end{document}